\documentclass[12pt]{article}

\usepackage[utf8]{inputenc}
\usepackage[T1]{fontenc}

\usepackage[margin=1in]{geometry}

\usepackage{amsmath}
\usepackage{amsfonts}
\usepackage{amssymb}
\usepackage{float}
\usepackage[normalem]{ulem}
\usepackage{booktabs}
\usepackage{tabularx}
\usepackage{threeparttable}
\usepackage{siunitx}
\usepackage{url}
\usepackage{graphics}
\usepackage[colorlinks=true, citecolor=blue, linkcolor=blue, filecolor=blue,urlcolor=blue,pdfauthor=author]{hyperref}

\usepackage[gen]{eurosym}

\graphicspath{{figures/}}

\usepackage{tikz}
\usepackage[europeanresistors,americaninductors]{circuitikz}
\usepackage{adjustbox}
\usepackage[shortcuts]{extdash}
\usepackage[colorinlistoftodos,prependcaption,textsize=tiny]{todonotes}
\usepackage{array}
\usepackage{tabularx}
\usepackage{tabulary}
\usepackage{subfigure}
\usepackage{graphicx}
\usepackage[font=small]{caption}
\usepackage{titlesec}
\titleformat{\section}
  {\normalfont\large\bfseries}{\thesection}{1em}{}
\titleformat{\subsection}
  {\normalfont\large\bfseries}{\thesubsection}{1em}{}
\titleformat{\subsubsection}
  {\normalfont\large\bfseries}{\thesubsubsection}{1em}{}
\usepackage{caption}
\usepackage[affil-it]{authblk}

\renewenvironment{abstract}
 {\noindent\bfseries\abstractname\normalfont}
 {\par\noindent\hrulefill\par\vspace{0.5cm}}

\begin{document}
%% \linenumbers
% \begin{frontmatter}

\title{Benefits from Islanding Green Hydrogen Production}
\author[1]{Christoph~Tries\thanks{Corresponding author: christoph.tries@tu-berlin.de}}
\author[1]{Fabian~Hofmann}
\author[1]{Tom~Brown}
\affil[1]{Technical University of Berlin, Department for Digital Transformation in Energy Systems, Einsteinufer 25a, 10587 Berlin, Germany}
\date{}

\maketitle

\begin{abstract}
In wind- and solar-dominated energy systems it has been assumed that there are synergies between producing electricity and electrolytic hydrogen since electrolysis can use excess electricity that would otherwise be curtailed. However, it remains unclear whether these synergies hold true at higher levels of hydrogen demand and how they compare with benefits of off\=/grid, islanded hydrogen production, such as better renewable resources and cost savings on electronics due to relaxed power quality standards. Using a mathematical model across two geographical locations for Germany, Spain, Australia, and Great Britain, we explore trade\=/offs and synergies between integrated and islanded electrolysers. Below a certain threshold, between \label{}5\% and \label{}40\% hydrogen share depending on the country, integrated electrolysers offer synergies in flexibility and reduced curtailment. Above these thresholds, islanded electrolysers become more favourable. Without cost advantages, systems including islanded electrolysers in Germany achieve up to \label{}21\% lower hydrogen costs than systems with only integrated electrolysers. With 25\% island cost advantage, this benefit rises to 40\% lower hydrogen costs. Our study identifies three investment regimes with country\=/specific transition points that vary based on island cost advantages and each country's renewable resources. Based on our results we provide guidelines for countries considering how to deploy electrolysers.
\end{abstract}

% \begin{keyword}
%   %% keywords here, in the form: keyword \sep keyword
%   %% Here are some suggestions:
%   renewable energy policy \sep hydrogen economy \sep large-scale integration of renewable power generation \sep electrolysers \sep system flexibility \sep islanded hydrogen production

%   %% PACS codes here, in the form: \PACS code \sep code

%   %% MSC codes here, in the form: \MSC code \sep code
%   %% or \MSC[2008] code \sep code (2000 is the default)
% \end{keyword}

% \end{frontmatter}

\hypertarget{introduction}{%
  \section{Introduction}\label{sec:introduction}}

% key role of h2
A key technological part of the transition to net\=/zero economies will likely be the wide\=/spread availability of green hydrogen\footnote{Green hydrogen is hydrogen produced by water electrolysis using renewable electricity.} and hydrogen\=/based fuels and feedstocks~\cite{rosenProspectsHydrogenEnergy2016,squadritoGreenHydrogenRevolution2023}. In the power sector, green hydrogen offers a solution for long\=/term energy storage, while in industries like steel and chemicals, it plays functional roles as a reducing agent and key material, respectively. Additionally, hydrogen derivatives have the potential to be a sustainable fuel option for shipping and aviation. This wide\=/ranging applicability makes hydrogen an important component in the gradual shift toward more sustainable energy systems and industrial practices. 
The anticipated rise in green hydrogen demand directly leads to the question of how countries position themselves in the emerging global market. Economic considerations, such as energy dependency, economic feasibility, and potential revenues from domestic markets, must be carefully weighed against each other to formulate optimal individual strategies. The European Union, for example, aims to produce 10 million tonnes of renewable hydrogen domestically by 2030, along with an equal amount of imports~\cite{europeancommission.directorategeneralforcommunication.REPowerEUActions2022,europeancommission.directorategeneralforcommunication.BoostingHydrogenEuropean2023}. Other countries like Chile or Australia aim to export large amounts of green hydrogen, which will likely exceed the total domestic energy demand by a large amount~\cite{StateWillPress2022,karOverviewHydrogenEconomy2023}. Furthermore, the ramp\=/up of the green hydrogen economy will likely unfold at a very high pace~\cite{odenwellerProbabilisticFeasibilitySpace2022}. As countries develop strategies to meet the quickly growing demand for green hydrogen, the question of optimal production methods thus becomes increasingly important in the near future. System planners and policy makers must carefully decide where to locate electrolysers~\cite{rabieeTechnicalBarriersHarnessing2021} and how to integrate them into the existing energy system.

% renewable system studies with integrated h2 production
The implementation of green hydrogen production integrated in power systems and sector\=/coupled energy systems has been extensively discussed in the scientific literature~\cite{mikovitsStrongerTogetherMultiannual2021,neumannPotentialRoleHydrogen2023,al-ghussainTechnoeconomicFeasibilityHybrid2023}. It is commonly acknowledged that integrated co\=/production of electricity and electrolytic hydrogen can benefit renewable energy systems by reducing curtailment of renewable power generation and providing a long\=/term storage solution~\cite{brownSynergiesSectorCoupling2018,rugglesOpportunitiesFlexibleElectricity2021}. To determine the optimal placement of integrated electrolysers, various consideration have to be taken into account. Emerging electrolysers are either strategically located near hydrogen demand centers and industrial clusters where demand exists~\cite{zhuIntegratedElectricityHydrogen2023}, or in resource\=/rich locations where electricity can be provided at low cost. This issue is not least related to the feasibility of expanding the existing power grid and/or creating a new hydrogen transport systems. In this regard, the study by Neumann et al.~\cite{neumannPotentialRoleHydrogen2023} looks at the impact of a hydrogen transport system in a fully renewable European energy system. The study shows that flexibility in transporting large quantities of hydrogen across the continent could reduce costs by up to 30\%, especially at industrial sites.

In contrast to integrated hydrogen production, the concept of dedicated hydrogen production sites, where all the electricity from dedicated wind and solar plants is used in electrolysis and which we refer to as ``hydrogen islands'', is increasingly being discussed in the literature. Various studies indicate the feasibility of hydrogen production in specialised systems like solar and wind hybrid setups~\cite{badeaHydrogenProductionUsing2017,galvanExportingSunshinePlanning2022,garciag.TechnicalEconomicCO22023,mazzeoGreenHydrogenProduction2022,panchenkoProspectsProductionGreen2023}. Specifically, the study by Garcia G. et al. demonstrated the economic feasibility of green hydrogen production in Chile~\cite{garciag.TechnicalEconomicCO22023} with prices ranging from US\$2.09/kg to US\$3.28/kg. Against this backdrop, ambitious large-scale green electrolysis projects are being planned in resource\=/rich countries such as Australia, Namibia, Morocco, and Chile, often as stand\=/alone systems without grid connections. The concept of ``energy islands'' is also gaining traction in the North Sea, where hydrogen is planned to be produced in remote areas and transported via pipelines to the mainland~\cite{luthHowConnectEnergy2023}. Studies like the one by Semeraro et al.~\cite{semeraroRenewableEnergyTransport2021}, have shown that hydrogen pipelines connecting hydrogen production centers are a cost-competitive method for the transmission of variable renewable energy compared to a high voltage direct current transmission lines.

While some studies like in~\cite{bhandariHydrogenEnergyCarrier2021} point out the flexibility advantages of grid-connected hydrogen production, islanding the production of hydrogen without connection to the electricity grid offers several techno\=/economic advantages:

\begin{enumerate}
\item Avoiding a public grid connection saves costs and potential connection delays.
\item Flexible choice of geographic sites abundant in renewable resources and away from public grids.
\item Simplified Balance of Plant (BoP) systems, leading to savings on expensive electronic components like inverters and rectifiers~\cite{beagleFuelingTransitionAccelerating2021}.
\item Lower power quality requirements for solar and wind power production.
\item Clear and unambiguous labeling as ``green hydrogen''.
\item The possibility to avoid conversion to alternating current and to use directly the DC current supplied by solar and wind facilities~\cite{liuDesignSuperconductingDC2018}.
\end{enumerate}

Thus far no research has explored the contrasting costs and benefits of islanded versus integrated hydrogen systems. Our study aims to address this gap in the literature. We employ optimisation techniques to evaluate the deployment of integrated and islanded electrolysers in fully renewable energy systems in individual countries. By contrasting optimal configurations with baseline scenarios---in which either integrated or island electrolysers are built exclusively---we describe characteristic investment regimes tailored to different hydrogen requirements. In addition, we consider the potential capital cost reductions in islanded systems due to simplified BoP configurations. For clarity and focus, our model has been streamlined to (1) consider only wind and solar energy sources, (2) limit the spatial locations of renewable resources to two per country, and (3) use stylised demand profiles for electricity and hydrogen.

The paper is structured as follows: Section~\ref{sec:methods} describes the model setup and the model runs. Section~\ref{sec:results} presents the results of our model runs. Section~\ref{sec:discussion} discusses the results and their implications, followed by section~\ref{sec:limitations} pointing out limitations. Finally, section~\ref{sec:conclusion} concludes the research.

\hypertarget{methods}{%
  \section{Methods}\label{sec:methods}}

\begin{figure}[!ht]
  \centering
  \includegraphics[width=0.7\textwidth]{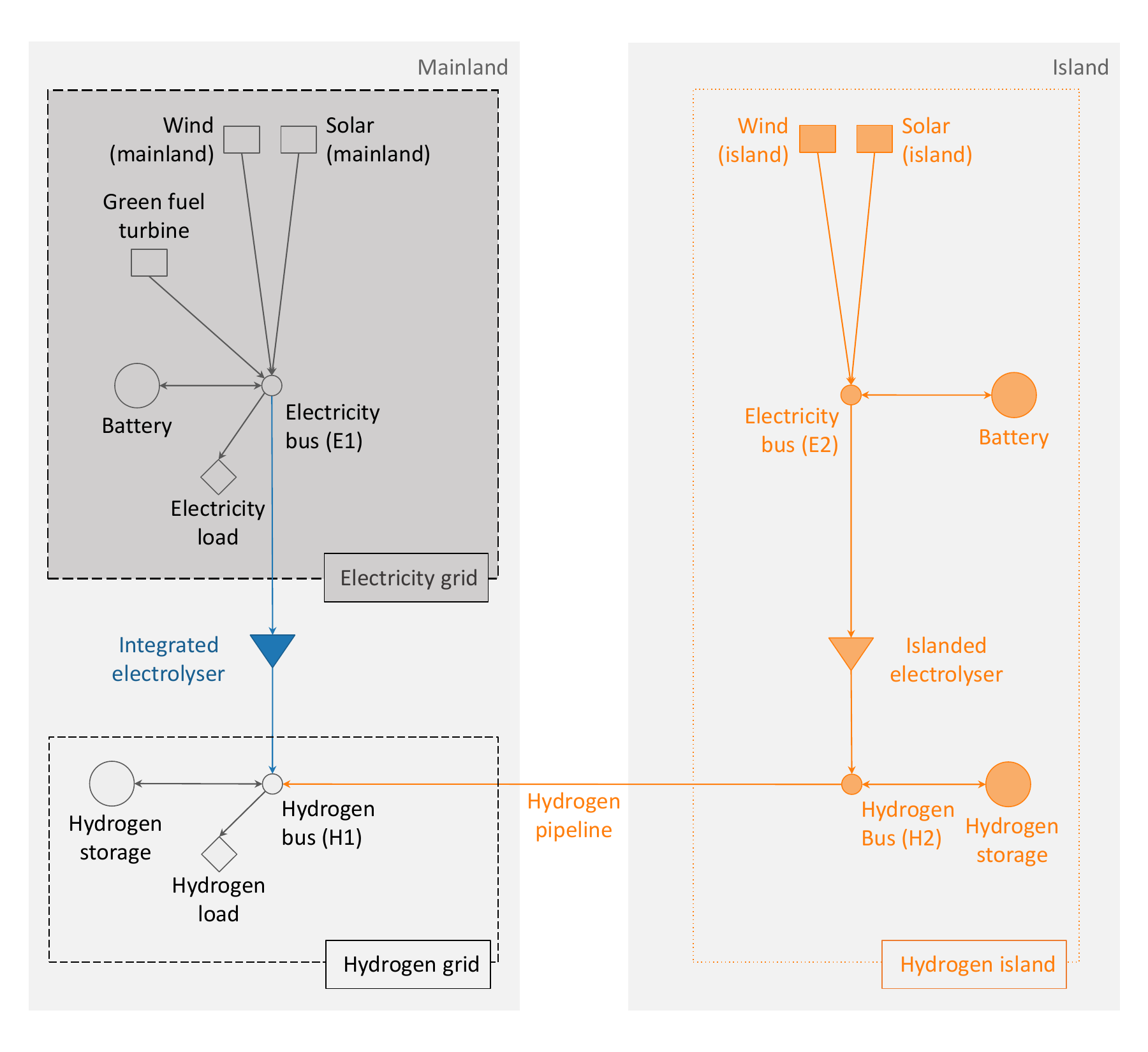}
  \caption{System setup for each scenario used throughout this study: Integration (grey and blue elements), Island (grey and orange elements) and Optimisation (all elements). The two loads are located on the mainland. The hydrogen demand is met by the mainland and/or island hydrogen production, the electricity demand is met by the mainland grid only.}
  \label{fig:systemSetupAll}
\end{figure}

We employ a streamlined model using the open\=/source toolbox PyPSA \cite{brownSynergiesSectorCoupling2018a} to analyse joint electricity and hydrogen production in individual countries. The model performs greenfield capacity expansion and hourly dispatch in systems based on wind and solar power, aiming to minimise annualised system costs. Each country is represented as two geographical locations with a fixed electricity and hydrogen demand (see Figure~\ref{fig:systemSetupAll}). The model incorporates wind and solar power, electrolysers, green fuel\=/powered dispatchable plants, short\=/term batteries, and large\=/scale underground hydrogen storage. Country\=/specific weather data for wind and solar generation are prepared using the open\=/source package ``atlite'' \cite{hofmannAtliteLightweightPython2021}. Technology costs projected for the year 2050 are retrieved from the Danish Energy Agency \cite{danishenergyagencyTechnologyDataGeneration2019} (see Table~\ref{tab:techInputData} for a complete list of assumptions). The model is available online as open\=/source software \cite{browntomBuildYourOwn}.

We select four countries to analyse the effects of available renewable resources and hydrogen demand on hydrogen costs, choice of electrolyser deployment, as well as renewables curtailment and electrolyser capacity factors. We make our country selection to include both countries with renewable systems dominated by wind and by solar power, with onshore and offshore island locations, and to include countries with island locations at longer and shorter distances from the mainland electricity grid. An overview of the selected countries is available in Table~\ref{tab:countries}.
%We chose these countries to represent a wide range of climatic conditions, renewable energy resources, and electricity systems.

\begingroup
\renewcommand{\arraystretch}{1.2}% Default value: 1
\newcolumntype{b}{>{\hsize=1.2\hsize}X}
\newcolumntype{s}{>{\hsize=.8\hsize}X}

\begin{table}[h!]
  \fontsize{8}{10}\selectfont
  \begin{center}
    \caption{Overview of considered countries and their main characteristics.}
    \label{tab:countries}
    \setlength\tabcolsep{4pt}
    \begin{tabularx}{\textwidth}{|>{\raggedright\arraybackslash}l|>{\raggedright\arraybackslash}s|>{\raggedright\arraybackslash}b|>{\raggedright\arraybackslash}X|>{\raggedright\arraybackslash}X|>{\raggedright\arraybackslash}X|>{\raggedright\arraybackslash}X|}
      \hline
      Country        & Distance to island [km] & Dominant mainland resource & Capacity factor wind (mainland) & Capacity factor wind (island) & Capacity factor solar (mainland) & Capacity factor solar (island) \\
      \hline
      United Kingdom & 269                     & wind                       & 0.40                            & 0.58                          & 0.11                             & -                              \\
      Spain          & 438                     & solar                      & 0.17                            & 0.24                          & 0.17                             & 0.18                           \\
      Germany        & 460                     & wind                       & 0.28                            & 0.63                          & 0.12                             & -                              \\
      Australia      & 1783                    & solar                      & 0.27                            & 0.25                          & 0.17                             & 0.18                           \\
      \hline
    \end{tabularx}
  \end{center}
\end{table}
\endgroup

\hypertarget{model-setup}{%
  \subsection{Model Setup}\label{sec:modelSetup}}

We model each country's energy system with three interconnected subsystems: one mainland electricity grid, one mainland hydrogen grid, and one hydrogen island (see Figure~\ref{fig:systemSetupAll}).

The mainland electricity grid serves electricity demand and can also power integrated electrolysers. Energy storage options on the mainland (batteries and hydrogen storage) and green fuel\=/powered combined cycle turbines help to align variable renewable production with demand profiles. Hourly capacity factors for wind
and solar represent spatially aggregated values with weightings prioritising sites with good renewable resources, i.e.
assuming renewable resources are distributed within a country proportionally to the square of the average capacity
factor.

The hydrogen island houses ``islanded'' electrolysers, which draw electricity only from dedicated renewable resources. Hydrogen produced on the island can be transported to the mainland via pipelines, with costs calculated based on capacity and distance (see Table~\ref{tab:transmissionInputData}). Slightly differing model designs were chosen to reflect country\=/specific circumstances: For Germany and Great Britain, offshore locations are chosen in the North Sea, where only offshore wind power is available. In Australia and Spain, pipelines are assumed to be above ground, whereas in Germany and Great Britain, they are modeled to be submarine with slightly higher costs, because of the offshore island locations and onshore portions running through densely populated areas.

The hydrogen pipeline connects hydrogen production with storage and consumption. It is designed to meet the hourly hydrogen demands either directly or by drawing from stockpiled hydrogen. All hydrogen must originally come from the mainland electricity grid or the hydrogen island.

\begin{figure}[h!]
  \centering
  % \begin{minipage}[position][height][inner-pos]{width}
  \begin{minipage}[t][4cm]{\textwidth}
  % \begin{minipage}[t][10cm]{9cm}
    \subfigure[Spain]{\label{fig:mapES}\includegraphics[height=3.6cm]{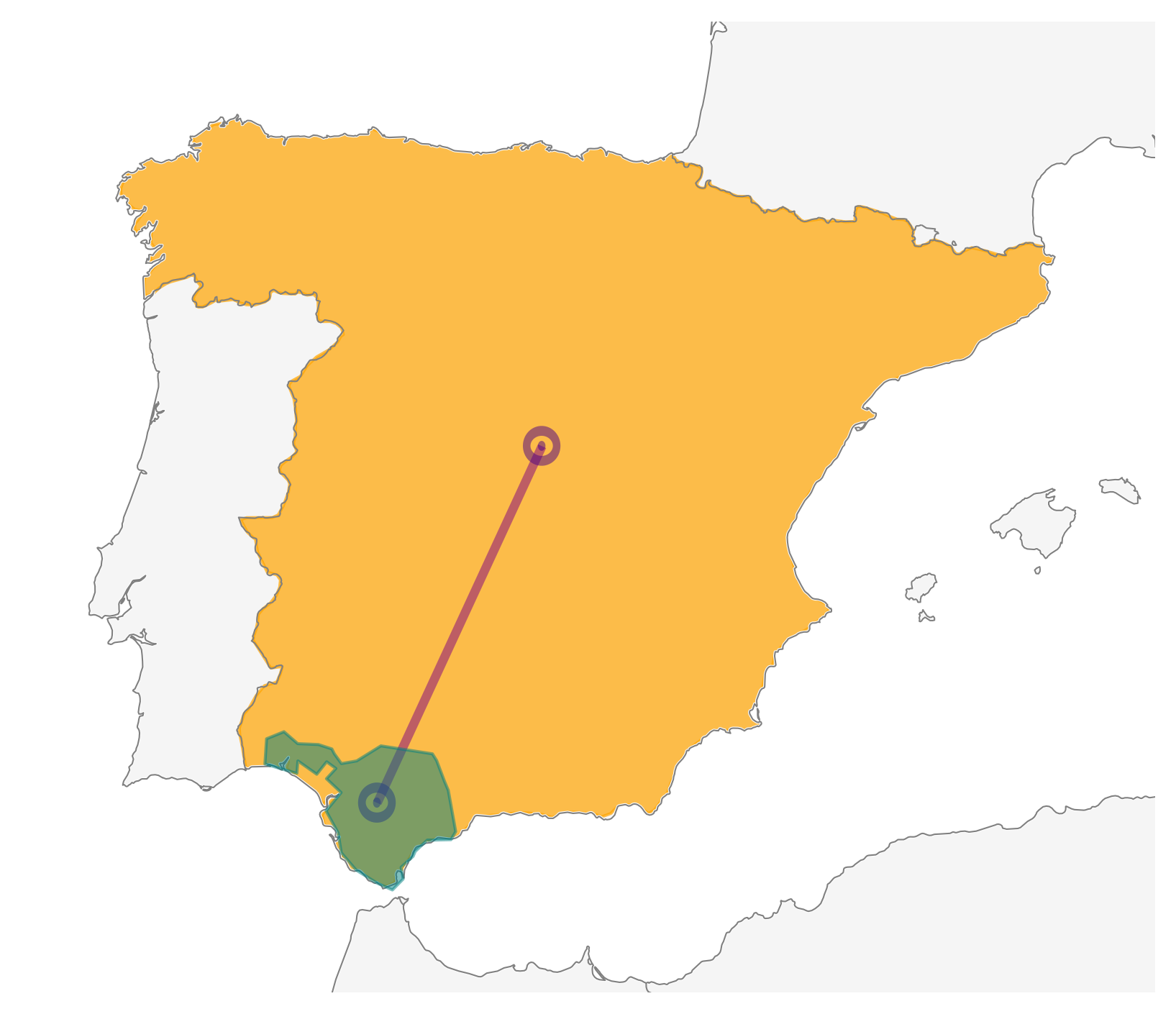}}
    \quad
    \subfigure[United Kingdom]{\label{fig:mapGB}\includegraphics[height=3.7cm]{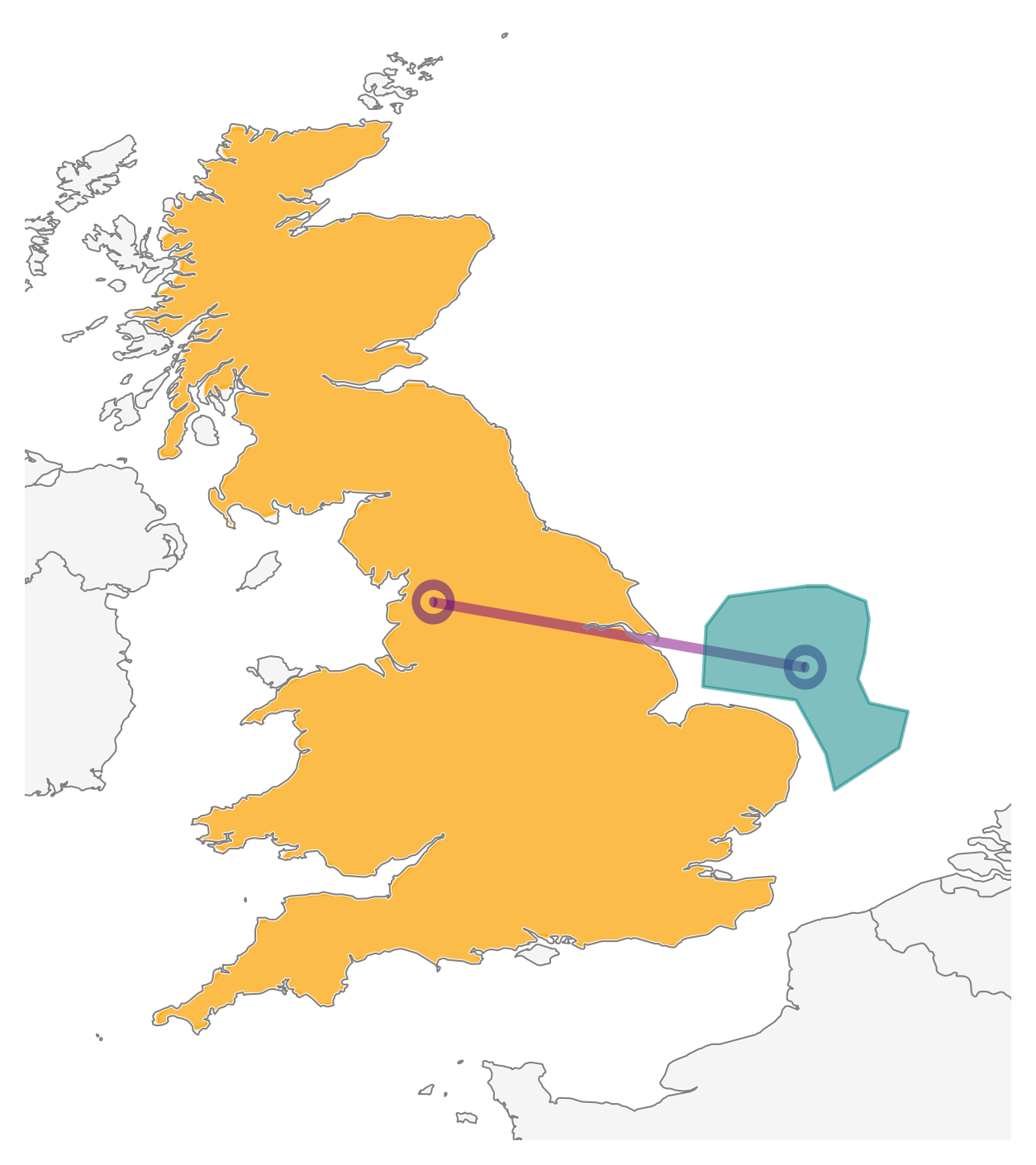}}
    \quad
    \subfigure[Germany]{\label{fig:mapDE}\includegraphics[height=3.6cm]{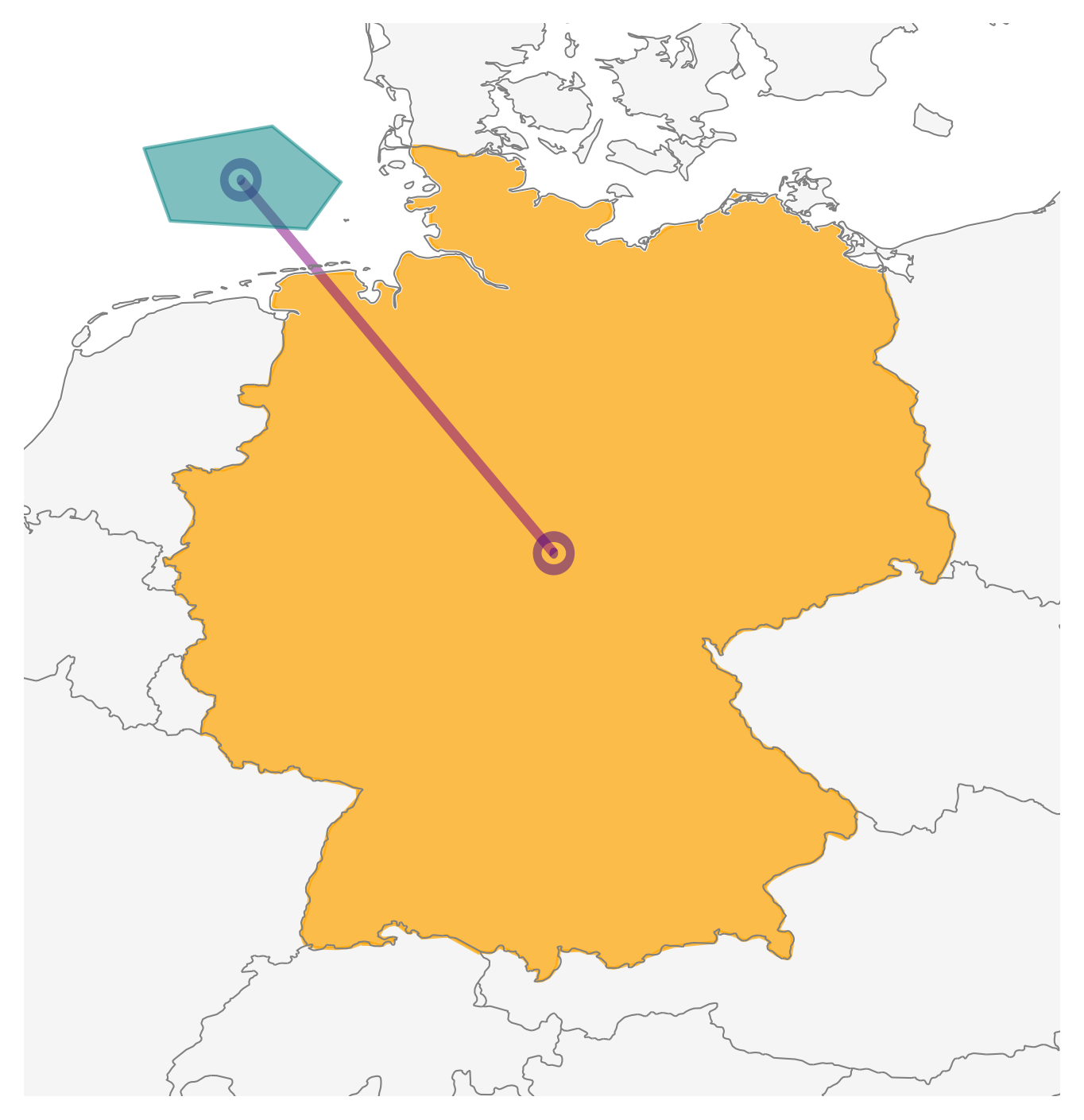}}
    \quad
    \subfigure[Australia]{\label{fig:mapAU}\includegraphics[height=3.6cm]{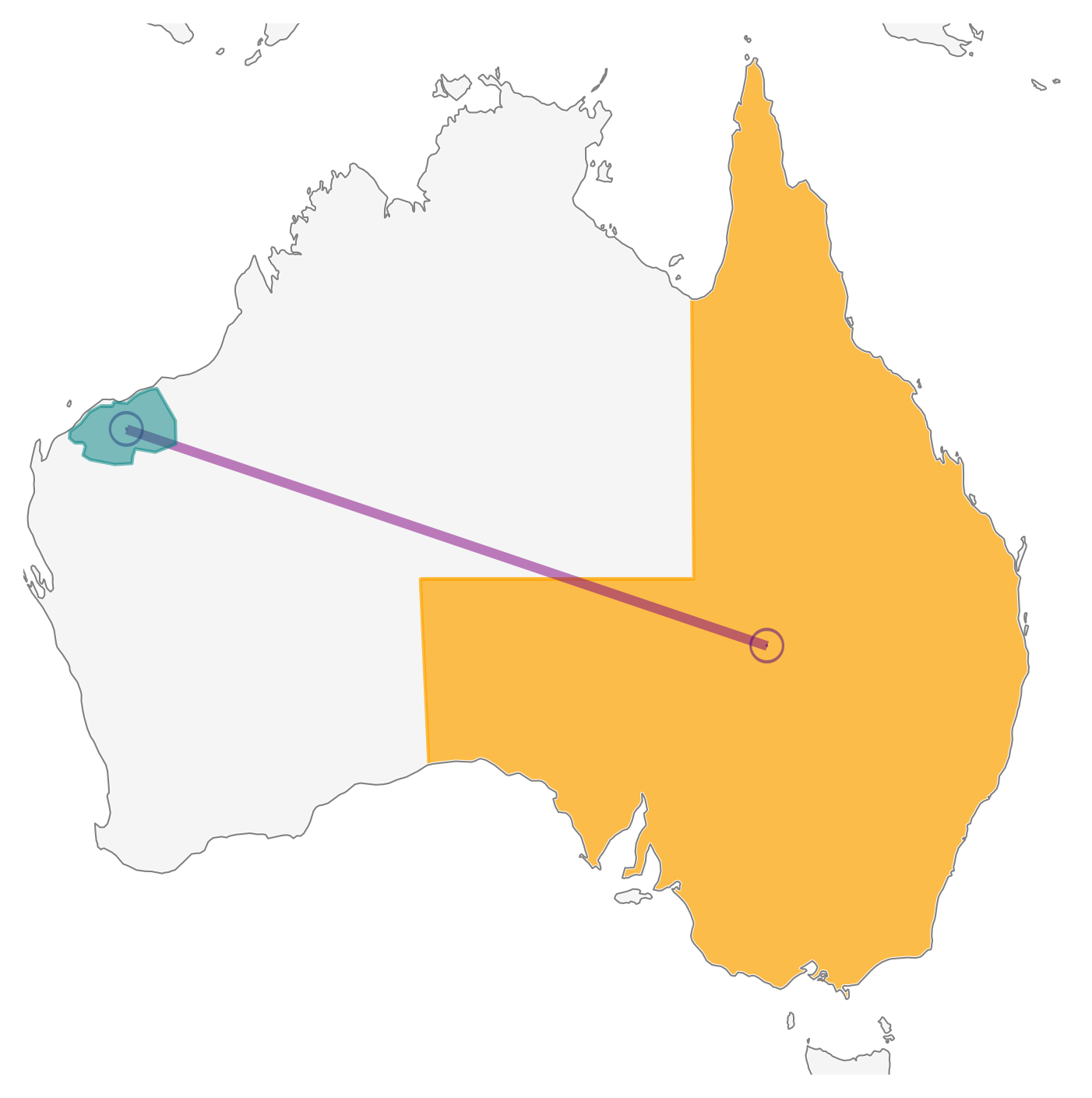}}
  % \end{minipage}
  \end{minipage}

  % \captionsetup{skip=10pt}
  \vspace{10pt}
  \caption{Geographical locations for the mainland (yellow) and island (blue) locations for each country considered in this study. The islands are located at sites with particularly good potential for renewable energy. Time\=/series for wind and solar power generation were obtained for each location using the atlite software~\cite{hofmannAtliteLightweightPython2021}.}
  \label{fig:maps}
\end{figure}

\hypertarget{model-runs}{%
  \subsection{Model Runs}\label{sec:modelRuns}}

We run our model for each country individually, and in three scenarios: (1) Integration: electrolysers are only built in the mainland electricity grid. (2) Island: electrolysers are only built on the island. (3) Optimisation: electrolysers can both be built in an integrated or in an islanded setting. The Optimisation scenario can utilise the advantages of both other scenarios, and its cost results thus represent a lower bound for the other two scenarios.

Rather than use the actual or projected demand for 2050, we instead explore the space of possibilities by sweeping the relative shares of hydrogen and electricity in final energy demand, going from 100\% electricity to 100\% hydrogen. We assume the demand for each energy carrier is constant throughout the year, and we use a single hourly wind and solar time series for each location, aggregated over its area.

As discussed above, producing hydrogen in an islanded setting can potentially provide cost benefits through simplified BoP equipment. We capture this aspect in the model by applying a uniform capital cost advantage to all power electronics elements in the islanded sub\=/system: capital costs for wind and solar power, electrolysers as well as the inverters used for battery storage. We then run each country, scenario, and hydrogen demand with six levels of these BoP\=/related capital cost advantages, ranging from 0\% (no advantage for the islanded system) to 25\% (see Table~\ref{tab:modelDim}).

\begin{table}[h!]
  \fontsize{8}{10}\selectfont
  \begin{center}
    \caption{Overview of dimensions for 1.800 model runs performed throughout this study.}
    \label{tab:modelDim}
    \setlength\tabcolsep{4pt}
    \begin{tabulary}{\textwidth}{|l|L|l|}
      \hline
      Dimension & Count of variables & Variables \\
      \hline
      Country & 4 & Spain, Germany, United Kingdom, Australia \\
      Electrolyser setting & 3 & Integration, Island, Optimisation \\
      Hydrogen share of final energy demand & 25 & from 0\% to 100\%\\
      Level of BoP-related capital cost reduction & 6 & 0\% to 25\%, in increments of 5\%\\
      \hline
    \end{tabulary}
  \end{center}
\end{table}

% \begin{table}[h!]
%   \fontsize{8}{10}\selectfont
%   \begin{center}
%     {\small \caption{Overview of dimensions for 1.800 model runs performed throughout this study.}}
%     \label{tab:modelDim}
%     \setlength\tabcolsep{4pt}
%     \begin{tabulary}{\textwidth}{|l|L|l|}
%       \hline
%       Dimension & Count of variables & Variables \\
%       \hline
%       Country & 4 & Spain, Germany, United Kingdom, Australia \\
%       Electrolyser setting & 3 & Integration, Island, Optimisation \\
%       Hydrogen share of final energy demand & 25 & from 0\% to 100\%\\
%       Level of BoP-related capital cost reduction & 6 & 0\% to 25\%, in increments of 5\%\\
%       \hline
%     \end{tabulary}
%   \end{center}
% \end{table}

\subsection{Cost-Optimal Electrolyser Investment Regimes}

\begin{figure}
  \centering
  \includegraphics[width=0.6\linewidth]{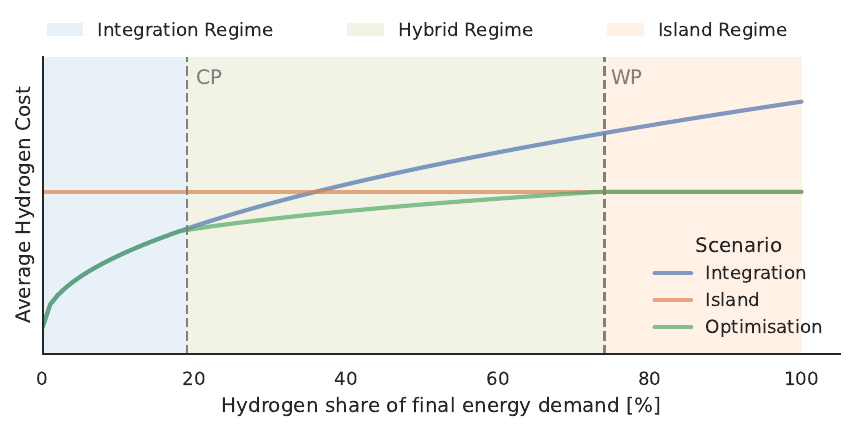}
  \caption{Idealised hydrogen cost curves for all three scenarios as a function of the hydrogen share of final energy demand. The curves show three investment regimes for the Optimisation scenario: The Integration Regime (left), where electrolysers are optimally deployed on the mainland; the Hybrid Regime (middle), where additional electrolysers are deployed both on the mainland and on the island; and the Island Regime (right), where all additional electrolysers are deployed on the island.}
  \label{fig:stylized-regimes}
\end{figure}

As discussed in the following sections, the optimal deployment of mainland and island electrolysers can be classified into characteristic patterns for each country. Figure~\ref{fig:stylized-regimes} sketches an idealised example of hydrogen production cost curves to showcase the classification of three optimal electrolyser investment strategies to support the analysis of our model results. We also identify two benefits from islanding electrolysers that determine the switch between these investment strategies, and we quantify the effect of these benefits on reducing hydrogen costs.

The idealised cost curves show the average hydrogen production costs as a function of the hydrogen\=/to\=/electricity share of final energy demand for the three different scenarios: Integration, Island, and Optimisation. The costs of hydrogen on the island are unaffected by the electricity production on the mainland and thus remain constant. The costs of hydrogen in the Integration Scenario rise, as the availability of otherwise\=/curtailed electricity decreases. Due to its higher degrees of freedom, the Optimisation scenario's costs are always equal to or below the costs of the Integration and Island scenarios.

The range of hydrogen demand is divided into three sections. In the first section, the hydrogen costs in the Optimisation scenario (green curve) follow the Integration scenario (blue curve). All electrolysers in the Optimisation system are deployed on the mainland, and no investments on the island are taken. At the ``Competitiveness Point'', the curves split and the ``Hybrid Regime'' starts. Here, islanded electrolysers become competitive with integrated electrolysers and it is cost\=/optimal to deploy electrolysers at the mainland as well as on the island. The deployment of both types of electrolysers in the Optimisation scenario leads to lower hydrogen costs than the two scenarios with all electrolysers deployed at one location (Integration and Island scenarios). Finally, at the ``Winout Point'' the curves of the Optimisation and the Island scenarios converge and the third regime starts. Islanded electrolysers are the cheapest source of additional hydrogen, and thus the Island Regime is characterised by the deployment of all additional electrolysers on the island.

\hypertarget{results}{%
  \section{Results}\label{sec:results}}

In this section, we explore the economics of hydrogen production for the selected set of countries. We delve into the key drivers that influence the cost\=/effectiveness of both integrated and islanded electrolysers. We identify the progression of investment regimes that dictate each country's electrolyser deployment behaviour. Lastly, we investigate how reductions in capital costs for islanded systems can alter these optimal electrolyser investment strategies.

\begin{figure}[h]
  \centering
  \includegraphics[width=\linewidth]{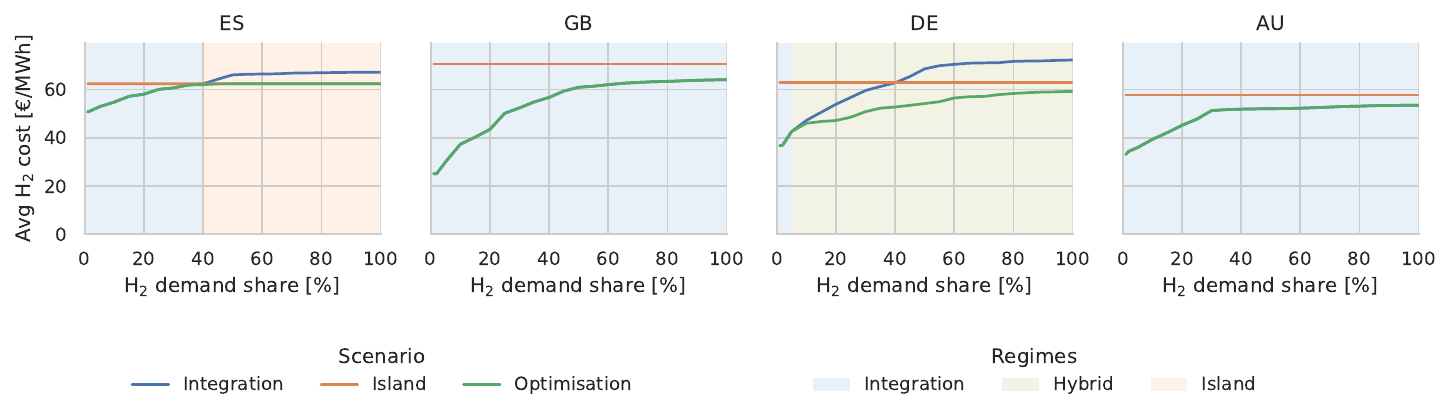}
  \caption{Average hydrogen costs for all countries in all three scenarios as a function of the hydrogen share of final energy demand. The hydrogen costs in the Optimisation scenario (green) are always equal to or lower than the hydrogen costs of the Integration (blue) and Island (orange) scenarios. In contrast to Spain and Germany, where the system reaches the Island and Hybrid Regime, respectively, for Great Britain and Australia the deployment of only integrated electrolysers (Integration Regime) is cost-optimal for all shares of hydrogen demand.}
\label{fig:avgH2Costs}
\end{figure}

Figure~\ref{fig:avgH2Costs} displays the average hydrogen costs for each country and scenario as a function of the hydrogen\=/to\=/electricity ratio in final energy demand. By model design, the hydrogen costs for the Island scenario (orange) remain constant across different hydrogen demand shares and are determined by the quality of the renewable resources at the island and the pipeline costs. Between the countries, hydrogen costs for the Island scenario vary from \label{res:avgH2-minmax-0-Isl}58€/MWh in Australia to 71€/MWh in Great Britain. In contrast, the hydrogen costs in the Integration scenario (blue) increase with hydrogen share, starting at their lowest values with 0\% hydrogen share, initially rising steeply before levelling off at round 30 to 50\% hydrogen share, asymptoting towards the cost of hydrogen at 100\% hydrogen share.

We now compare the Island and Integration scenarios. At 100\% hydrogen share, the Island scenario has lower costs than the Integration scenario for Spain and Germany. In both countries, the two scenarios reach cost equality at 40\% hydrogen share. For Australia, despite superior island resources (see Table~\ref{tab:countries}), pipeline costs render hydrogen production on the island more expensive even for 100\% hydrogen share. In Great Britain, the island scenario has better wind resources but lacks solar power, making the average hydrogen costs on the island more expensive than on the mainland. Pipeline costs also contribute to a small degree but are not solely responsible for the cost difference.

Finally, the hydrogen costs in the Optimisation scenario (green) are by design always equal to or below the hydrogen costs of the Integration and Island scenarios. Consequently, for Great Britain and Australia, the Optimisation scenario always deploys the same system as the Integration scenario. For these two countries, the system stays in the Integration Regime and it is always cost\=/optimal to invest into integrated electrolysers. In contrast, the optimal deployment pattern for electrolysers in Spain and Germany reaches a tipping point: For Spain, the system moves at 40\% hydrogen share from the Integration Regime directly into the Island Regime, where it is cost\=/optimal to deploy all additional electrolysers on the island. For Germany, the system enters the Hybrid Regime at around 5\% hydrogen share, where both mainland and island electrolysers are deployed, and remains in this regime until 100\% hydrogen share. 

In the following, we delve into the unique characteristics of each regime, we identify two benefits that make islanded electrolysers more cost\=/effective than integrated electrolysers, and we quantify the impact of these benefits on the average hydrogen costs for each country.

\subsection{Integration Regime - No Benefits for Islanded Electrolysers}\label{sec:integrationRegime}

First, we analyse the Integration Regime which is characterised by the fact that only integrated electrolysers are deployed to accommodate the rising demand for hydrogen. The Integration Regime is prevalent in Spain from 0 to 40\%, in Germany from around 0 to 5\% hydrogen share, and in Great Britain and Australia for the entire range of hydrogen demand shares.

Two advantages underpin the economic viability of integrated electrolysers in the Integration Regime. First, for countries like Australia and Great Britain, mainland\=/based hydrogen production with integrated electrolysers is more cost\=/effective than island production regardless of the hydrogen demand level (see Figure~\ref{fig:avgH2Costs}). But second, even if islanded electrolysers are more cost\=/effective than integrated electrolysers on their own as is the case for Spain and Germany, the co\=/production of hydrogen and electricity on the mainland offers synergies that reduce both hydrogen and electricity production costs. 
% These co\=/production synergies are particularly seen at lower hydrogen demand levels and they are the reason that average hydrogen costs in the Integration and Optimisation scenarios slope upwards with increasing hydrogen shares. 
These co\=/production synergies mostly accrue early for smaller shares of hydrogen, with average hydrogen costs quickly converging to the cost of electrolysis with no co\=/production of electricity (see hydrogen costs for the Integration scenario in Figure~\ref{fig:avgH2Costs}).

\begin{figure}[h]
  \centering
  \includegraphics[width=\linewidth]{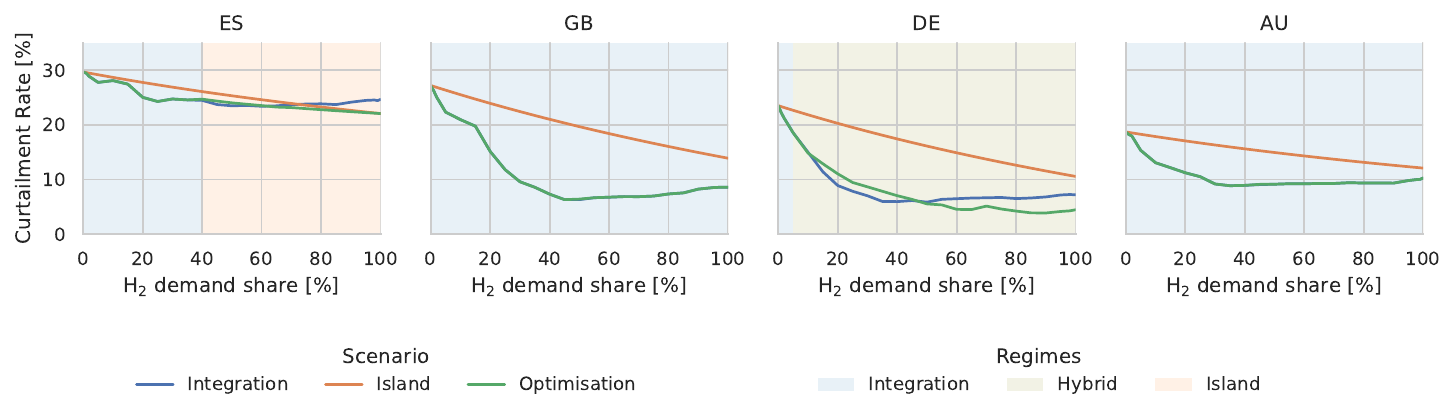}
  \caption{Average Curtailment Rate for countries in all three scenarios as a function of the hydrogen share of final energy demand. The Curtailment Rate is defined as the ratio of curtailed energy to total energy demand. Across all countries and scenarios, the electrolyser deployment in the Optimisation scenario (green) reduces the curtailment rate. The extent of reduction is determined by synergies between electricity and hydrogen production on the mainland, and the quality of renewable resources at the mainland and the island.}
  \label{fig:curtailment}
\end{figure}

The root of this synergy is the utilisation of previously curtailed energy for hydrogen production. Evidence for this is the decrease in curtailment quantities in the Integration Regime for the Integration and Optimisation scenarios, as highlighted in Figure~\ref{fig:curtailment}. However, it is essential to note that electrolysers are not capable of utilising all the available curtailed energy. Each country exhibits one cost\=/optimal level of curtailment for electricity production (see 0\% hydrogen share in Figure~\ref{fig:curtailment}) and a different cost\=/optimal level of curtailment for hydrogen production (see 100\% hydrogen share in Figure~\ref{fig:curtailment}). Both levels of curtailment are determined by each country's renewable resources. The difference between these optimal curtailment levels delineates the quantity of curtailed energy that can feasibly be directed toward electrolysers as hydrogen demand increases. For example, whilst Spain experiences high levels of curtailment---up to \label{}30\% of all generated electricity---only about \label{}8 percentage points of this curtailment are mitigated through the deployment of integrated electrolysers.

\subsection{Hybrid Regime - Benefit from Synergies between Mainland and Island Resources}\label{sec:hybridRegime}

\begin{figure}[h]
  \centering
  \includegraphics[width=\linewidth]{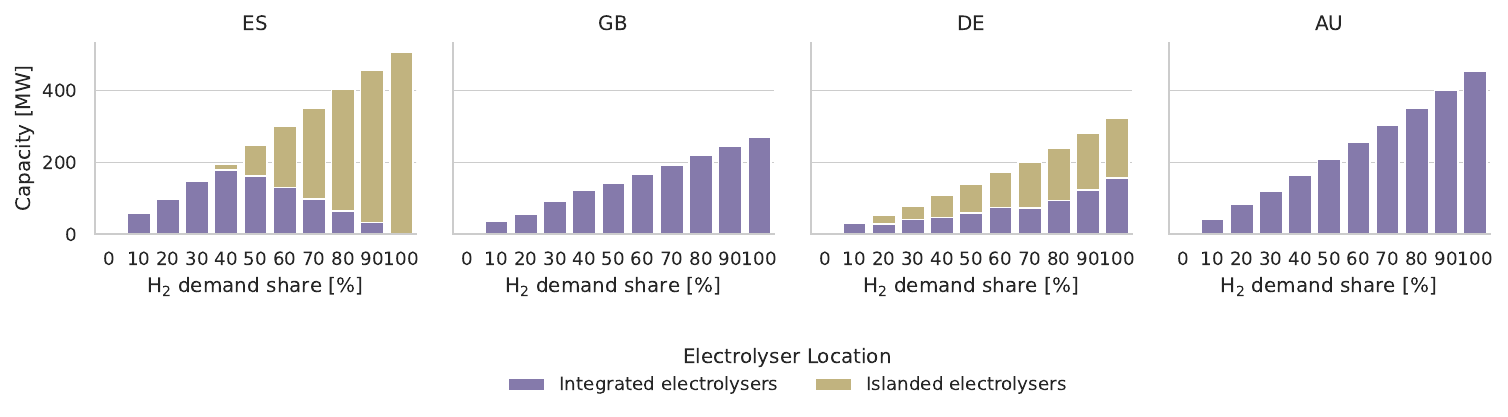}
  \caption{Installed electrolyser capacities for all countries in the Optimisation scenario as a function of the hydrogen share of final energy demand. In the first regime (Integration Regime), the system only deploys integrated electrolysers (purple). Once the system moves to the Hybrid Regime, as is the case in Germany at 10\%, integrated and islanded electrolysers (yellow) are used in parallel. The Island Regime, which starts in Spain at 40\%, is characterised by additional electrolysers being deployed only on the island.}
  \label{fig:installedElyCapacity}
\end{figure}

The Hybrid Regime is characterised by the deployment of electrolysers both on the mainland and on the island to meet additional hydrogen demand. This regime is prevalent in Germany from 10 to 100\% hydrogen share (see Figure~\ref{fig:installedElyCapacity}). The Integration Regime gives way at the ``Competitiveness Point'' (CP), where islanded electrolysers become cost\=/competitive with integrated electrolysers for the first time. It can also be identified as the hydrogen share where the first islanded electrolyser is deployed in each country (see Figure~\ref{fig:installedElyCapacity}).

Islanded electrolysers become competitive in the Hybrid Regime because they exploit synergies between the different renewable resource profiles on the mainland and on the island to produce hydrogen in both locations. We call this the Portfolio Benefit.

\begin{figure}[h]
  \centering
  \includegraphics[width=.7\linewidth]{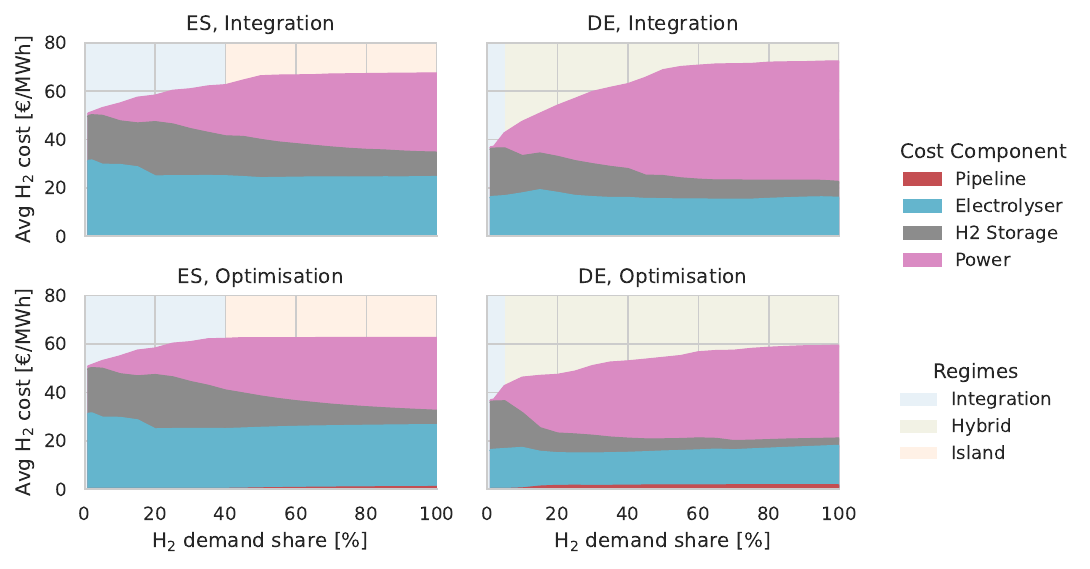}
  \caption{Hydrogen cost breakdown by cost components for the Integration (top row) and Optimisation (bottom row) scenarios for Spain and Germany. Within the Integration Regime, the two scenarios result in the same setup and cost breakdown. While the cost breakdown does not differ much in the Island regime (left side, Spain), a strong difference can be seen in the Hybrid regime (right side, Germany). Here, the Optimisation scenario benefits from synergies between the mainland and the island, which reduce the need for hydrogen storage and decrease electrolyser investment costs.}
  \label{fig:h2CostArea}
\end{figure}

The synergies responsible for the Portfolio Benefit stem from two sources: The first is a reduced requirement for hydrogen storage, as corroborated by a comparison of the specific hydrogen storage cost components between the Integration and Optimisation scenarios for Germany, depicted in Figure~\ref{fig:h2CostArea}. The time series of renewable energy production at both locations align in such a way that hydrogen production schedules throughout the year complement each other, thereby minimising the need for hydrogen storage.

\begin{figure}[h]
  \centering
  \includegraphics[width=.8\linewidth]{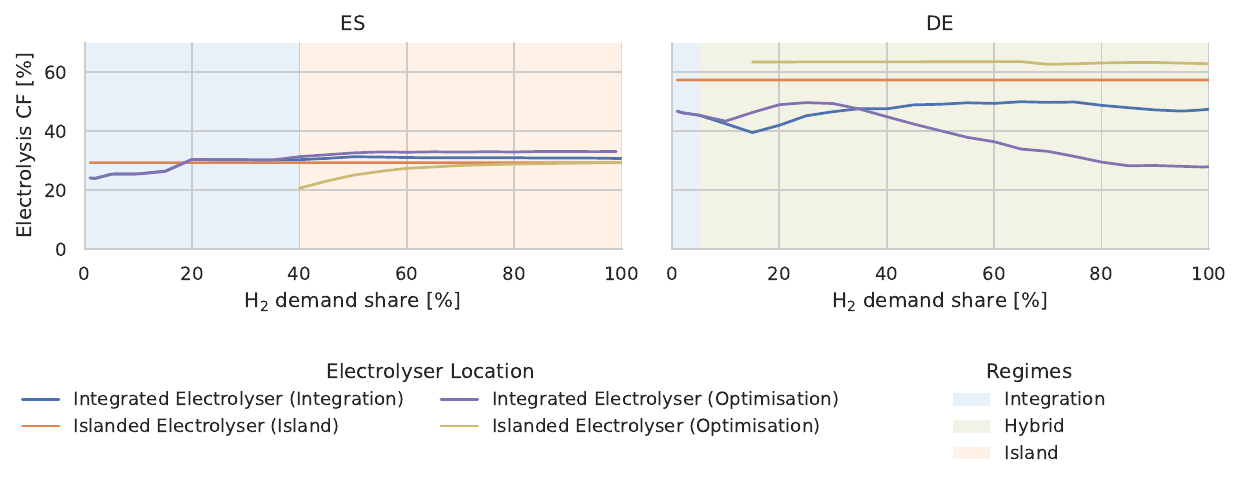}
  \caption{Electrolysis capacity factors for Spain and Germany for integrated and islanded electrolysers in all scenarios. Whereas the capacity factors in Spain stay relatively constant, the capacity factors for integrated electrolysers in the Optimisation scenario (purple) in Germany decrease with the hydrogen share, in favour of maximum islanded hydrogen production (yellow).}
  \label{fig:cfElectrolysis}
\end{figure}

The second is lower specific electrolyser investment costs (see also  Figure~\ref{fig:h2CostArea}). In the Optimisation scenario, Germany deploys electrolysers on the island which run 100\% of the time that wind power is available. This leads to a capacity factor of 65\% for islanded electrolysers compared to about 45\% for integrated electrolysers in the Integration scenario (see Figure~\ref{fig:cfElectrolysis}). Islanded electrolysers operate at full capacity (or at least whenever wind energy is available), whilst integrated electrolysers selectively operate during periods of zero\=/cost (i.e., using curtailed energy) or low\=/cost electricity and/or when their operation can lead to hydrogen storage reduction. The capacity factor for integrated electrolysers remains also around 45\% such that the average capacity factor for the entire electrolyser fleet in the Optimisation scenario is larger than in the Integration scenario (at least for the range from 10 to 60\% hydrogen share).

The value of the Portfolio Benefit is measured as the difference between the hydrogen costs of the Integration and Optimisation scenarios while in the Hybrid Regime (see Figure~\ref{fig:avgH2Costs}). The only country where the Portfolio Benefit is active is Germany, and at the maximum it reduces hydrogen costs by \label{res:portfolio-DE-0}21\% (15€/MWh) compared to the Integration scenario.

Three findings already emerge from these observations. Firstly, even at 0\% hydrogen share, the marginal cost for the system to generate hydrogen is small but significant. Specific power costs for hydrogen are zero, but specific electrolyser investment costs and specific hydrogen storage costs still make up between \label{res:avgH2-minmax-0-Opt}25€/MWh (Spain) and 51€/MWh (Australia), depending on the country studied (see Figure~\ref{fig:h2CostArea}). Secondly, hydrogen costs from integrated electrolysers start low, but quickly rise as hydrogen demand increases (see Figure~\ref{fig:avgH2Costs}). And thirdly, when the resource portfolio from the mainland and island locations fits well together to produce hydrogen, the Portfolio Benefit reduces hydrogen costs for the Optimisation scenario compared to the Integration scenario (see Germany starting at 5\% hydrogen share in Figure~\ref{fig:avgH2Costs}).

\subsection{Island Regime - Benefit from Superior Island Resource Quality}\label{sec:islandRegime}

The Island Regime is defined as the regime under which additional hydrogen demand is solely met by deploying islanded electrolysers. Here, we see that the islanded electrolyser outcompetes the integrated electrolyser due to its better resources. We call this the Superiority Benefit for islanded electrolysers. The Superiority Benefit is also quantified as the difference between the hydrogen costs of the Integration and Optimisation scenarios, but only applies under the Island Regime (see Figure~\ref{fig:avgH2Costs}). The transition to the Island Regime occurs at the ``Winout Point'' (WP), and this is where the Superiority Benefit outweighs the Portfolio Benefit. Any integrated electrolysers deployed prior to reaching this regime remain in the system at a constant rate relative to the declining electricity demand as hydrogen demand increases.

The Island Regime is prevalent in Spain, accounting for the range from 40 to 100\% hydrogen share (see Figure~\ref{fig:avgH2Costs}). Spain does not produce any Portfolio Benefit for hydrogen production, since the time series for renewable resource availability in both locations (mainland and island) are very similar, with the island only possessing a slight advantage in capacity factor for solar energy production. Since there is no Portfolio Benefit, the CP and WP coincide and the system moves directly into the Island Regime. The Superiority Benefit exists only for Spain, starting at 40\% hydrogen share, and decreases hydrogen costs by up to \label{res:superiority-ES-0}7\% (5€/MWh) compared to the Integration scenario.

\hypertarget{increased-benefits-from-capital-cost-advantages}{%
  \subsection{Increased Benefits from Capital Cost Advantages to Islanded Electrolysers}\label{sec:costredux}}

\begin{figure}[h]
  \centering
  \includegraphics[width=\linewidth]{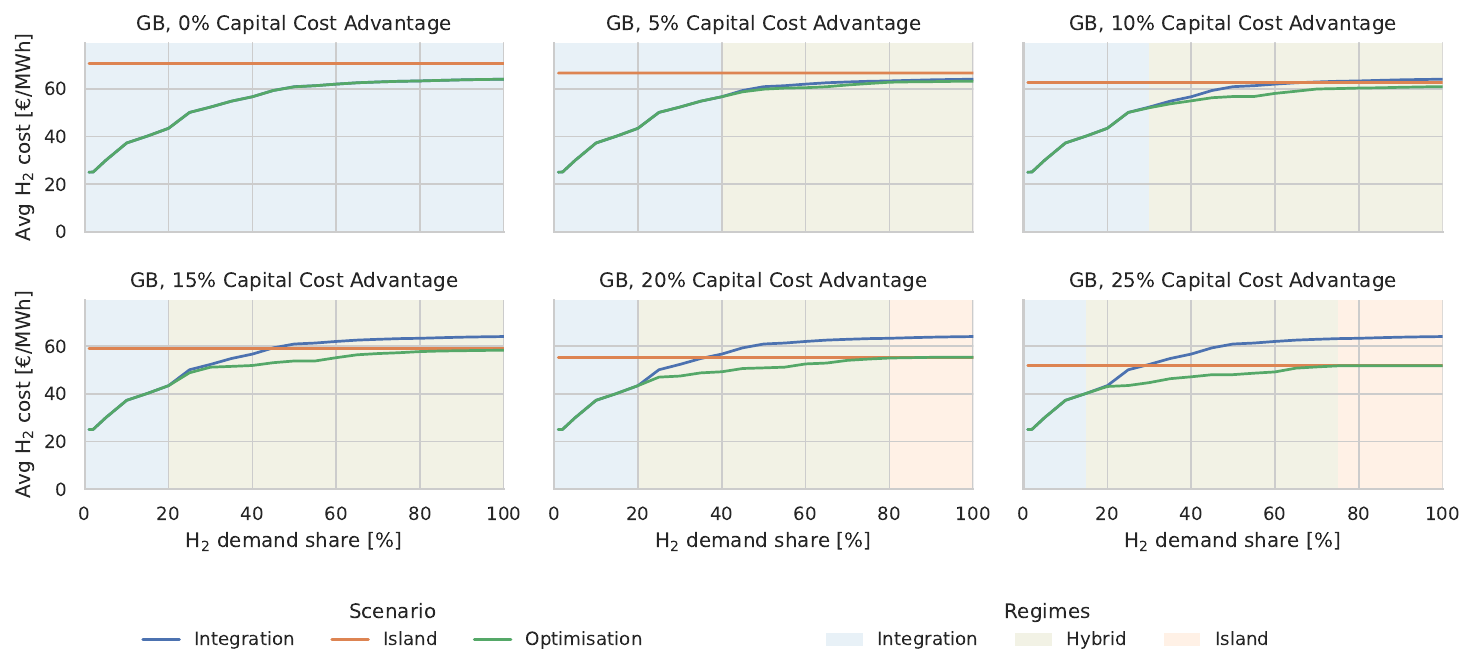}
  \caption{Hydrogen costs for Great Britain at 0\% to 25\% capital cost advantage for all three scenarios. The capacity cost advantages pull the costs of islanded hydrogen production down, and thus shift the start of the Hybrid Regime, and the start of the Island Regime to the left.}
  \label{fig:avgH2Costs-GB}
\end{figure}

In addition to the dynamics of competition between integrated and islanded electrolysers, our study also includes the influence of lower capacity costs, from hereon denoted as cost advantage, for relevant system components on the hydrogen island. On an island, power electronic costs could be saved because grid\=/level power quality no longer needs to be maintained, or the conversions to alternating current could be avoided altogether \cite{beagleFuelingTransitionAccelerating2021}. In this section, we exemplarily discuss the results for Great Britain (see Figure~\ref{fig:avgH2Costs-GB}).

We still observe the same dynamics, trade\=/offs and synergies between integrated and islanded electrolysers, but the locations of the regimes shift. In addition, incorporating cost advantage leads to lower hydrogen costs from islanded electrolysers (see decreasing level of orange curves in Figure~\ref{fig:avgH2Costs-GB}, as cost advantage increases). As a result for Great Britain, the CP and WP are gradually introduced (CP at 5\%, WP at 20\% cost advantage), and then shift left as cost advantage increases further. At 5\% cost advantage, the decrease in hydrogen costs from islanded electrolysers is not yet sufficient to outright compete with the synergies between integrated electrolysers and the power system. However, some small amount of Portfolio Benefit exists and barely makes the hybrid deployment of integrated and islanded electrolysers beneficial for the overall system.

This finding shows that even a modest decrease in capacity costs by only five percentage points can significantly change the electrolyser deployment regimes. As illustrated in Figure~\ref{fig:avgH2Costs-GB}, the cheaper islanded hydrogen production pushes the CP to the left, and at 10\% cost advantage thus introduces the Hybrid Regime at \label{}40\% hydrogen share. With increasing cost advantage, the CP and Hybrid Regime shift forward to \label{}15\% hydrogen share. At 20\% cost advantage, the WP is finally reached as well, and the system enters the Island Regime at 80\% hydrogen share. From hereon, the system deploys all additional electrolysers on the island.

The two benefits identified in our study also change with increasing cost advantage. For example, for Great Britain without cost advantage there is no Superiority Benefit to islanded electrolysers, because renewables on the mainland are actually the superior resource. With 25\% cost advantage, the island becomes the superior resource, and applies a benefit of up to \label{res:superiority-GB-25}19\% lower hydrogen costs (12€/MWh) to islanded electrolysers. Net shifts in favour of islanded electrolysers are similar for Australia with \label{res:superiority-AU-25} 20\% (11€/MWh) lower hydrogen costs, but even larger for Spain and Germany \label{res:superiority-ES-25}\label{res:superiority-DE-25}(33\% and 37\% lower costs, respectively, see Figure~\ref{fig:avgH2Costs-Grid}).

The Portfolio Benefit also increases significantly as cost advantage increases. For Great Britain, the maximum Portfolio Benefit at 5\% cost advantage amounts to \label{res:portfolio-GB-5}3\% (2€/MWh, at 65\% hydrogen share), and at 25\% cost advantage to \label{res:portfolio-GB-25}21\% (13€/MWh, at 50\% hydrogen share).

Finally, it is worth noting that the engineering\=/based cost advantages contribute to the same effect as the resource advantage of the island over the mainland. Essentially, an island location with strong and favourable renewable resources would exhibit the same boost to the competitiveness of islanded electrolysers as a less favourable location with reduced capital costs.

\hypertarget{discussion}{%
  \section{Discussion}\label{sec:discussion}}
These results have significant implications for countries to develop individually tailored policies for their future clean hydrogen procurement. Integrated hydrogen production is optimal at low levels of hydrogen demand. These integrated electrolysers provide flexibility to the renewable electricity system on the mainland and allow otherwise\=/curtailed electricity to be used. As the share of hydrogen in final energy demand and thus also the amount of integrated electrolysers on the mainland increase, Portfolio Benefits can make islanded electrolysers competitive. For countries with a high domestic hydrogen demand, and those looking to export hydrogen, the optimal system deployment can include the bulk of electrolysers operating in an islanded setting, especially if islanded hydrogen production yields some of the capital cost advantages studied in this paper. In the following, we briefly discuss these findings and focus on their implications for system planners as well as additional regulatory and technology issues. 

\begin{figure}[h]
  \centering
  \includegraphics[width=\linewidth]{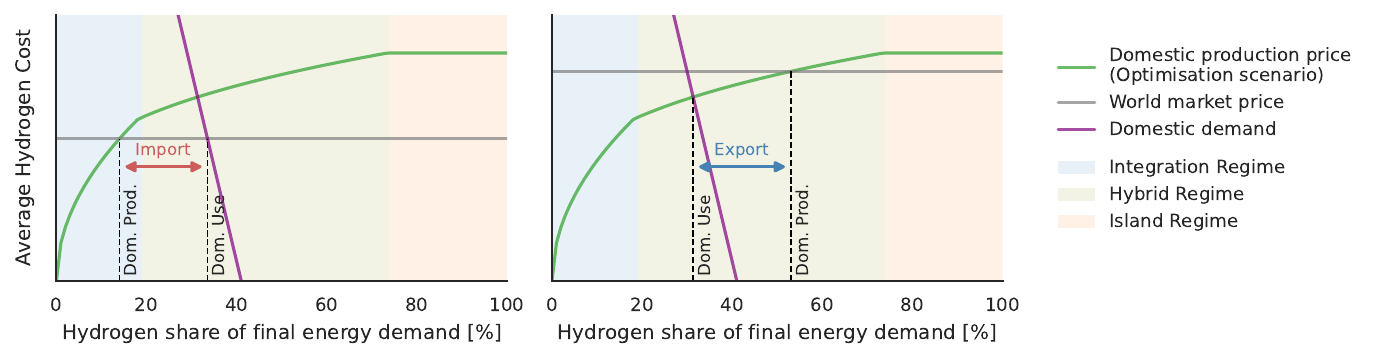}
  \caption{Idealised cost curves to determine domestic hydrogen production and import/export volumes, as well as deployment patterns for electrolysers. The intersection of the domestic hydrogen production costs from the Optimisation scenario (green) and the world market price (grey) determine the domestic hydrogen production level. The intersection of the domestic hydrogen demand and the lower of both price curves determine the level of domestic hydrogen use. The difference between domestic hydrogen production and hydrogen use is balanced by imports from or exports to the world market.}
  \label{fig:stylized-regimes-discussion}
\end{figure}

From a system planning perspective, two additional quantities have to be estimated in order to formulate an optimal investment strategy for integrated versus islanded electrolysers: the domestic hydrogen demand and the expected world market price for hydrogen. Assuming these as given, we show, on the basis of Figure~\ref{fig:stylized-regimes-discussion}, how to determine (1) the optimal amount of domestically produced hydrogen (and how much to import or export accordingly to meet domestic hydrogen use), (2) whether to deploy integrated or islanded electrolysers, and/or at which respective shares, and (3) how to unfold the electrolyser deployment over time as hydrogen demand ramps up.

% In section \ref{sec:dynamics}, we assess the impact of core exogeneous influence factors on the system dynamics and our main results. In section \ref{sec:benefits} we discuss the benefits that we have identified in our study, and finally in section \ref{sec:limitations} we address the limitations of our study and further work.

% \subsection{Tasks and Insights for System Planners}\label{sec:systemPlanners}
To determine the optimal amount of domestically produced hydrogen (1), system planners need to construct their domestic hydrogen production price curve based on the optimal investment regimes for integrated and islanded electrolysers, as we deduce with the Optimisation scenario in our paper. The intersection with the world market price then determines the optimal domestic hydrogen production level (see Figure~\ref{fig:stylized-regimes-discussion}).

The level of domestic hydrogen use can then be derived as the intersection of the domestic hydrogen demand curve and the lower bound of the domestic and world market hydrogen price curves. Import and export volumes then make up the difference between domestic hydrogen production and domestic hydrogen use, unless considerations like security of supply require a different procurement mix.

The mix of integrated and islanded electrolysers (2) can easily be deduced from Figure~\ref{fig:installedElyCapacity} which displays the electrolyser mix deployed in the Optimisation scenario (or from Figure~\ref{fig:installedElyCapacity-Grid} for cost advantages).

Finally, to determine the ramp up of electrolysers (3), system planners can interpret the x\=/axis of increasing hydrogen demand as a timeline. At very low levels of hydrogen in final energy demand, which is today's starting point for almost all countries, every country studied should start by deploying integrated electrolysers to take advantage of the benefits of co\=/production with electricity (the only exceptions in our study being Spain above 15\% cost advantage and Germany above 20\% cost advantage, where hydrogen from islanded electrolysers has always lower costs than from integrated electrolysers). The crucial decision points are then to know (a) when to start deploying islanded electrolysers alongside integrated electrolysers (i.e., when the system reaches the Competitiveness Point and enters the Hybrid Regime) and which mix of electrolysers to deploy, and (b) when to freeze the deployment of further integrated electrolysers on the mainland (i.e., when the system reaches the Winout Point and enters the Island Regime).

This picture is of course dependent on external factors, domestic hydrogen demand, the world market price as well as the optimised hydrogen use (Optimisation scenario). In particular, a higher (lower) domestic hydrogen demand leads to higher (lower) domestic hydrogen use and a corresponding increase in imports/decrease in exports (or vice versa). 
% It does not impact the optimal domestic hydrogen production levels. It does not impact the pattern of the optimal investment strategies.
On the other hand, a lower (higher) world market price for hydrogen results in a lower (higher) optimal domestic hydrogen production level, higher (lower) domestic hydrogen use, and thus leads to increased imports/decreased exports (or vice versa). 
% It has no impact on the pattern of the optimal electrolyser investment strategies, but by reducing (increasing) the optimal domestic hydrogen production level, it also reduces (increases) the amount of electrolysers deployed, and possibly moves the mix of electrolysers to include more integrated (islanded) electrolysers.
Finally, a lower (higher) hydrogen production cost in the Optimisation scenario leads to a shift of the CP, WP and the regimes to the left (right). It increases (decreases) the optimal domestic hydrogen production level at the same time, thus increasing the amount of electrolysers deployed, and changing the mix to include more islanded (integrated) electrolysers. Such a shift to lower (higher) hydrogen production costs could be achieved by higher (lower) cost advantages for islanded electrolysers or by superior (inferior) island resources.
Given these varying external factors, it becomes evident that a robust regulatory framework is essential for guiding investments and ensuring sustainable hydrogen production.

Furthermore, we show with our study that even at close to 0\% hydrogen share, the marginal cost for the system to generate hydrogen is not zero. We can see that all countries with at least some hydrogen demand deploy integrated electrolysers, thus reducing some, but not all of their previously curtailed renewables generation. This additional electricity demand for hydrogen production bolsters electricity prices in previously zero- or low\=/cost hours, leading to an increase in the average power cost components for hydrogen production with increasing hydrogen demand (also demonstrated by Ruhnau~\cite{ruhnauHowFlexibleElectricity2022}). The average electrolyser capacity factors remain significantly below 100\% and are closely tied to the renewable resource type with which they are co\=/located, confirming results from Cloete et al.~\cite{cloeteCapitalUtilizationHydrogen2021}.

From a regulatory standpoint, hydrogen islands provide an unquestionably green source of hydrogen. However, as we show in our study all countries should start by deploying at least \textit{some} integrated electrolysers. Also, other than our assumption of an energy system mainly based on wind and solar resources, in the real world there will be at least some carbon emissions remaining in the energy system, even as countries move forward on their way to net\=/zero economies. Therefore, developing a robust regulatory framework to ensure that (highly cost\=/effective) integrated electrolysers also ensure the production of \textit{green} hydrogen is of paramount importance. A further crucial regulatory consideration must be the identification and implementation of the correct incentives so that investments into electrolysers realised by the private sector follow (roughly) the optimal investment regimes.

Developing remote sites with dedicated renewables and islanded electrolysers also offers advantages when grid connections for additional renewables experience significant delays or soft cost barriers (e.g., permitting, local public opposition), which is a major challenge in many markets. Remote sites for renewables may not experience as much public opposition as centralised sites, and could therefore be developed more quickly.

Finally, on the technology side, as we show in our study the reduction of power electronic costs in islanded systems can yield significant benefits both to reduce hydrogen costs, and to make benefits from islanded resources accessible in hydrogen production at even lower levels of hydrogen demand. There is cost saving potential in operating islanded collector grids with lower\=/quality power, as long as they do not connect to the AC network. Using a pure DC power system would further reduce the number of electronic equipment parts needed for the BoP, especially expensive instruments like inverters and rectifiers, and also avoids conversion losses \cite{beagleFuelingTransitionAccelerating2021}. The power electronics for BoP systems make up around 35\% of the cost of polymer electrolyte membrane electrolysers, and significant cost\=/advantage potentials across the entire electrolyser system are possible in the near term \cite{beagleFuelingTransitionAccelerating2021}. Simplifying power electronics equipment for such islanded subsystems can thus make already attractive remote renewable resources even better.

\section{Limitations}\label{sec:limitations}
To get a clearer view on the dynamics, we have provided a simple modelling framework that enables a grounded analytical understanding of the factors and trade\=/offs at play in an energy system coupling electricity and hydrogen. However, the model does not include detailed transmission networks for electricity and hydrogen, and demand profiles are constant. For some of the island locations with smaller distance to the mainland, including the option of connecting the superior resource to the power system with HVDC cables could also strengthen the results. There is a strong dependence on the chosen areas and resulting resource profiles for the mainland and the island of each country. The reliance on only one resource per location is also a strong simplification. The island could also directly export to other markets, thus avoiding pipeline costs. Furthermore, the results have limited relevance for short\=/term decisions, since fossil fuels are ignored, and we focus on an energy system powered largely by variable renewable energy sources. Instead, the model is intended to provide long\=/term guidance about electrolyser deployment in hydrogen production. In further work, we aim to demonstrate the effects of this interplay in a more detailed networked model for Europe, PyPSA\=/Eur \cite{brownSynergiesSectorCoupling2018a}, which can address many of the shortcomings.

\hypertarget{conclusion}{%
  \section{Conclusion}\label{sec:conclusion}}

We have explored the trade\=/offs between integrating water electrolysis and electricity generation versus islanding hydrogen production. Thereby, we have identified three investment regimes: The Integration Regime, where only integrated electrolysers are deployed; the Hybrid Regime, where both integrated and islanded electrolysers are deployed; and the Island Regime, where only islanded electrolysers are deployed. Transitions between these regimes are determined by renewable resource quality and capital cost advantages. At low hydrogen demand, integrated electrolysers are generally favoured. However, as demand increases, systems transition to the Hybrid or Island Regime where mostly islanded electrolysers are deployed.

Furthermore, we have identified two types of benefits a system can derive from deploying islanded electrolysers: (1) Portfolio Benefits, which are synergies of locational co\=/production of hydrogen on the island and the mainland that reduce specific hydrogen storage and electrolyser investment costs, and (2) Superiority Benefits, a hydrogen production cost benefit derived from islanded electrolysers being connected to a superior renewable resource than integrated electrolysers.

Portfolio Benefits arise in the Hybrid Regime when a mix of integrated and islanded electrolysers is deployed. In Germany, they reduce hydrogen costs by up to \label{res:portfolio-DE-0}21\% (15€/MWh) without cost advantages, and by up to \label{res:portfolio-DE-25}40\% (28€/MWh) with cost advantages of 25\%. Superiority Benefits reduce hydrogen costs for islanded electrolysers under the Island Regime. At no cost advantages, Superiority Benefits exist for Spain and Germany, reaching up to \label{res:superiority-ES-0}7\% (5€/MWh) and \label{res:superiority-DE-0}18\% (13€/MWh) lower costs, respectively. With cost advantages of 25\%, the island is the superior resource location for all countries: hydrogen cost benefits range from \label{res:superiority-GB-25}19\% (12€/MWh) for Great Britain to \label{res:superiority-DE-25}37\% (27€/MWh) for Germany.

We provide several guidelines for system planners. If local hydrogen production is likely to be low, either because of low demand or inexpensive imports, countries should focus on integrated electrolysis. If however, large demand or export volumes are expected, there may be cost benefits from islanding production, particularly if renewable resources are better or cost reductions for islanded equipment are expected.

\section*{CRediT authorship contribution statement}

\textbf{Christoph Tries}: Conceptualization, Methodology, Software, Validation, Formal analysis, Investigation, Writing --- original draft, Writing --- review \& editing, Visualization. \textbf{Fabian Hofmann}: Conceptualization, Software, Writing --- review \& editing, Visualization. \textbf{Tom Brown}: Conceptualization, Software, Writing --- review \& editing, Supervision, Funding acquisition.

\section*{Acknowledgements}
Christoph Tries and Fabian Hofmann are funded by the Breakthrough Energy Project “Hydrogen Integration and Carbon Management in Energy System Models”.

\bibliographystyle{IEEEtran}

{\scriptsize\bibliography{references}}

\newpage

\section*{Appendix}

% Add prefix "A-" to figure and table numbers in Appendix
\renewcommand{\thefigure}{A\arabic{figure}}
\renewcommand{\thetable}{A\arabic{table}}

\begingroup
\setlength{\tabcolsep}{6pt}% Default value: 6pt
\renewcommand{\arraystretch}{1.2}% Default value: 1

\begin{table}[ht!]
  \fontsize{8}{10}\selectfont
  \begin{center}
    \caption{Technology input data for 2050 used as a basis for all countries and scenarios.}
    \label{tab:techInputData}
    \setlength\tabcolsep{4pt}
    \begin{tabulary}{\textwidth}{|L|L|L|L|L|L|L|L|L|}
      \hline
      & & & & & Battery & Battery & Hydrogen & \\[-3pt]% compensate for extrarowheight
      & Wind & Wind & & Electro- & storage & storage & storage & Green fuel \\[-3pt]% compensate for extrarowheight
      & (onshore) & (offshore) & Solar & lyser & (inverter) & (battery) & (cavern) & turbine \\
      \hline
      Capital cost [€/kW] & 960 & 1380 & 240 & 400 & 60 & 75 & 0.7 & 800 \\
      Fixed O\&M & & & & & & & & \\[-3pt]% compensate for extrarowheight
      [\% of capital cost] & 3 & 3 & 3 & 3 & 3 & 3 & 14 & 3 \\
      Fuel cost [€/MWh] & - & - & - & - & - & - & - & 100 \\
      Efficiency [\%] & - & - & - & 67 & 95 & - & - & 60 \\
      Lifetime [years] & 25 & 25 & 25 & 20 & 15 & 15 & 25 & 25\\
      \hline
    \end{tabulary}
  \end{center}
\end{table}
\endgroup

\begingroup
\setlength{\tabcolsep}{6pt}% Default value: 6pt
\renewcommand{\arraystretch}{1.2}% Default value: 1

\begin{table}[ht!]
  \fontsize{8}{10}\selectfont
  \begin{center}
    \caption{Hydrogen pipeline costs input data for 2050}
    \label{tab:transmissionInputData}
    \setlength\tabcolsep{4pt}
    \begin{tabularx}{.8\textwidth}{|>{\raggedright\arraybackslash}l|>{\raggedright\arraybackslash}X|>{\raggedright\arraybackslash}X|>{\raggedright\arraybackslash}X|>{\raggedright\arraybackslash}X|>{\raggedright\arraybackslash}X|>{\raggedright\arraybackslash}X|>{\raggedright\arraybackslash}X|}
      \hline
                           & H2 pipeline compressor & H2 pipeline (overground) & H2 pipeline (submarine) \\
      \hline
      Capital cost         &                        &                          &                         \\[-3pt]% compensate for extrarowheight
      [€/kW, €/kW/km]      & 4.48                   & 0.226                    & 0.329                   \\
      Fixed O\&M           &                        &                          &                         \\[-3pt]% compensate for extrarowheight
      [\% of capital cost] & 1.7                    & 1.5                      & 3                       \\
      Lifetime [years]     & 20                     & 50                       & 30                      \\
      \hline
    \end{tabularx}
  \end{center}
\end{table}
\endgroup

% \begin{figure}[!ht]
%   \centering
%   \subfigure[Integration]{\label{fig:systemSetupIntegration}\includegraphics[width=.33\textwidth]{../setup/hydrogen-island-diagram-v4(integrated).pdf}}
%   \subfigure[Island]{\label{fig:systemSetupIsland}\includegraphics[width=.33\textwidth]{../setup/hydrogen-island-diagram-v4(islanded).pdf}}
%   \subfigure[Optimisation]{\label{fig:systemSetupOptimisation}\includegraphics[width=.33\textwidth]{../setup/hydrogen-island-diagram-v4(all).pdf}}
%   \caption{System setups for each scenario}
%   \label{fig:systemSetupThree}
% \end{figure}

\begin{figure}[!ht]
  \centering
  \includegraphics[width=1.0\textwidth]{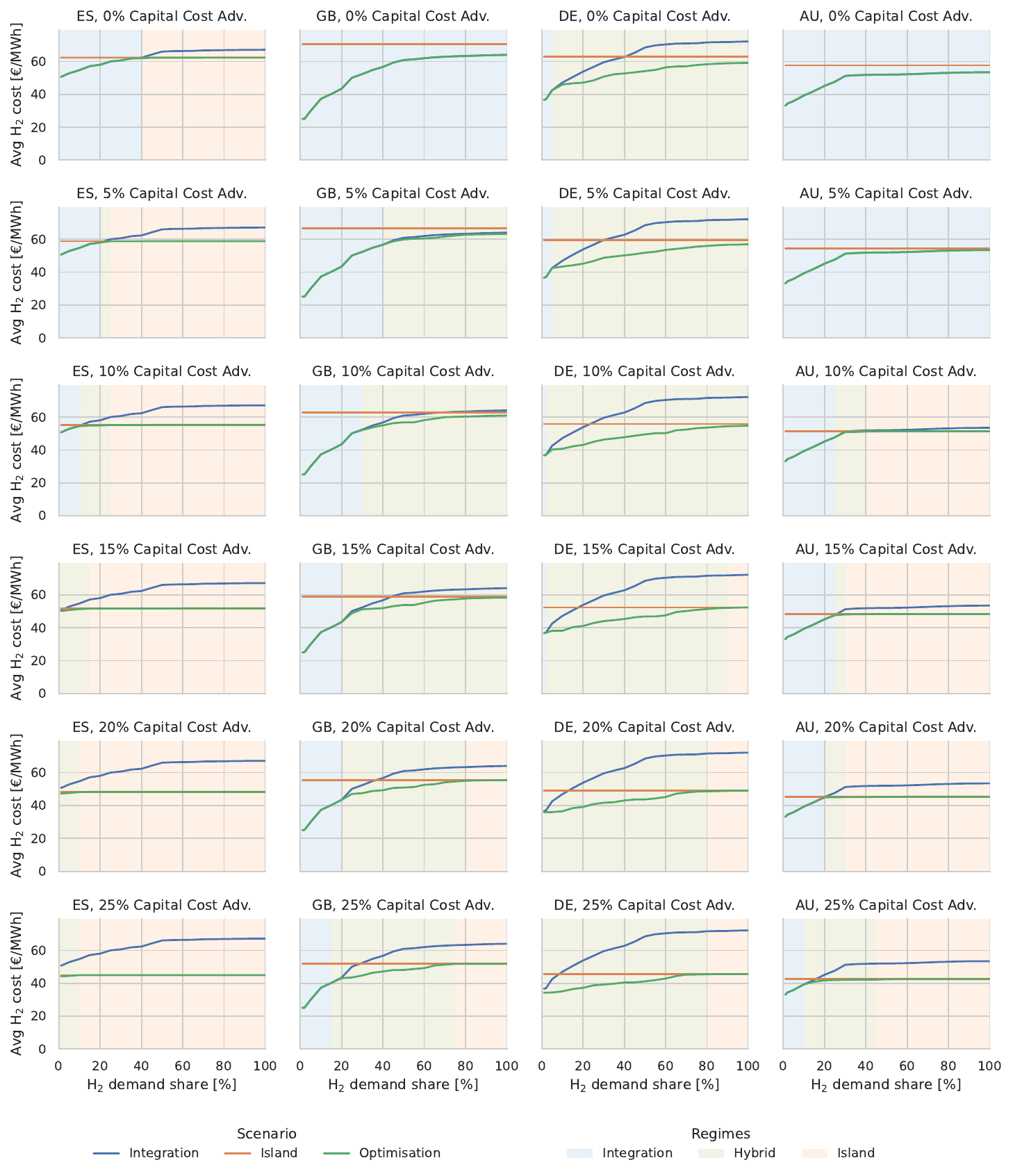}
  \caption{Average hydrogen prices across countries and levels of capital cost reductions}
  \label{fig:avgH2Costs-Grid}
\end{figure}

\begin{figure}[!ht]
  \centering
  \includegraphics[width=1.0\textwidth]{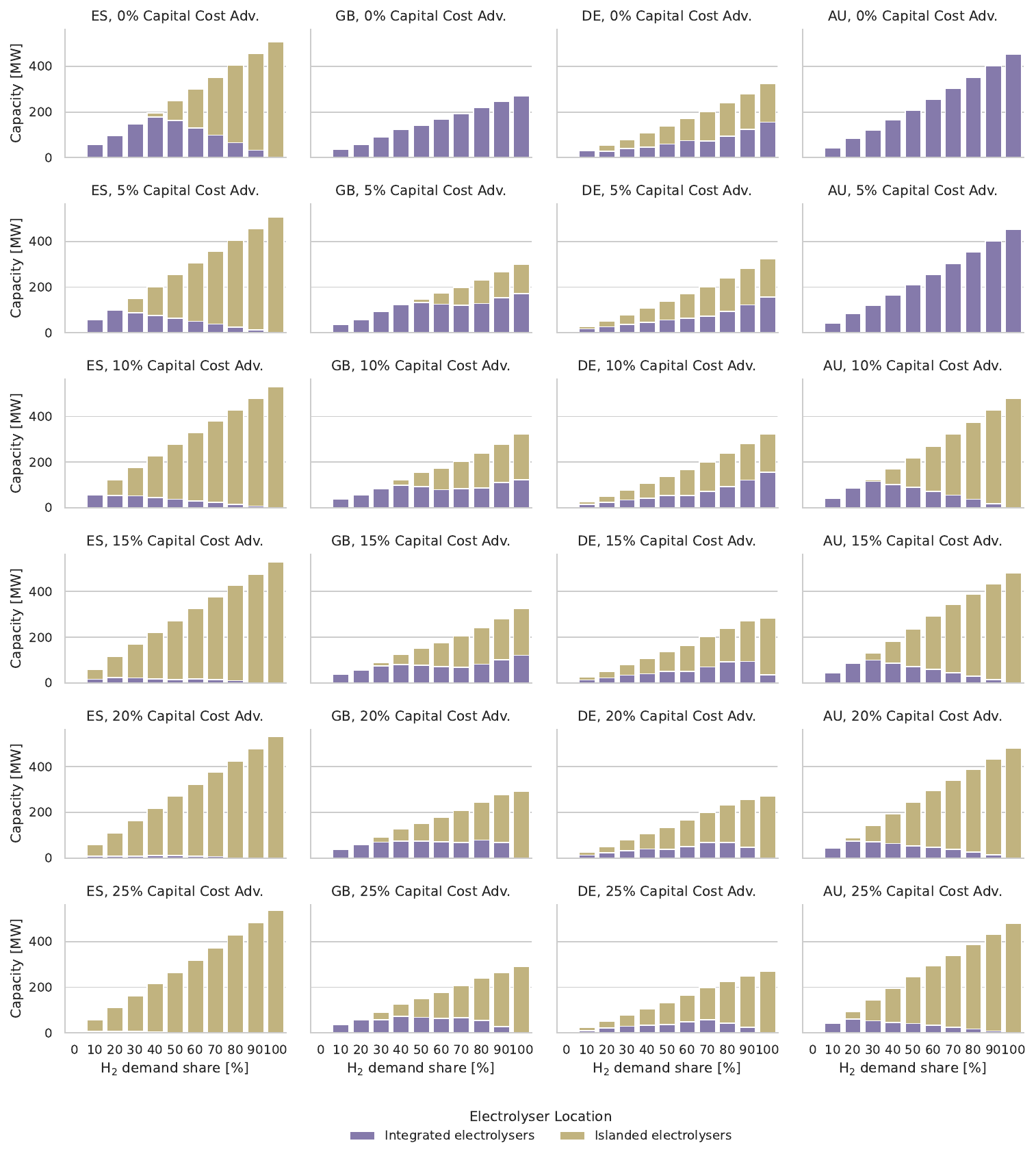}
  \caption{Installed electrolyser capacities for Optimisation scenario across countries and levels of capital cost reductions}
  \label{fig:installedElyCapacity-Grid}
\end{figure}

\begin{figure}[!ht]
  \centering
  \includegraphics[width=1.0\textwidth]{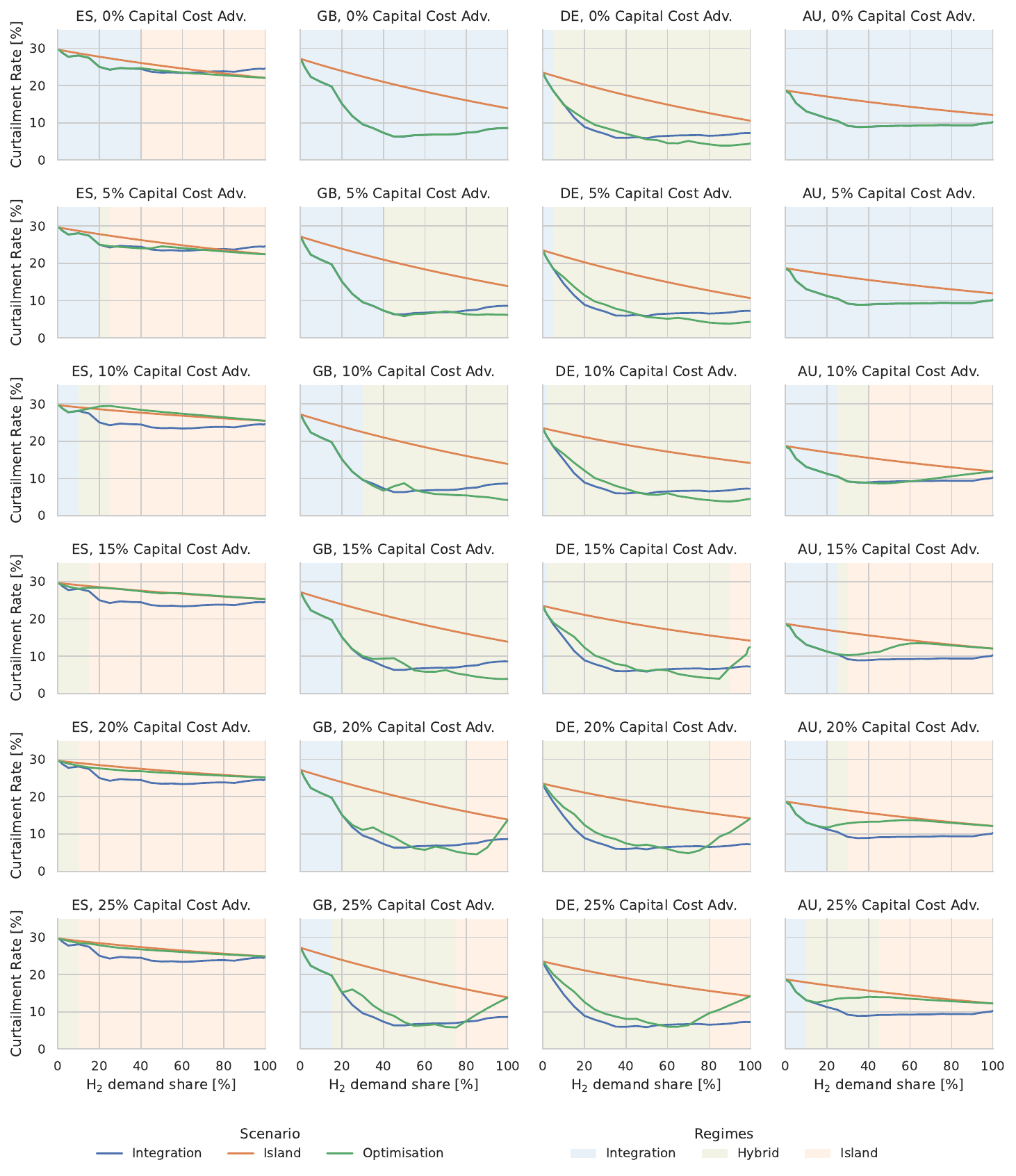}
  \caption{Curtailment rate for all scenarios across countries and levels of capital cost reductions}
  \label{fig:curtailment-Grid}
\end{figure}

\begin{figure}[!ht]
  \centering
  \includegraphics[width=1.0\textwidth]{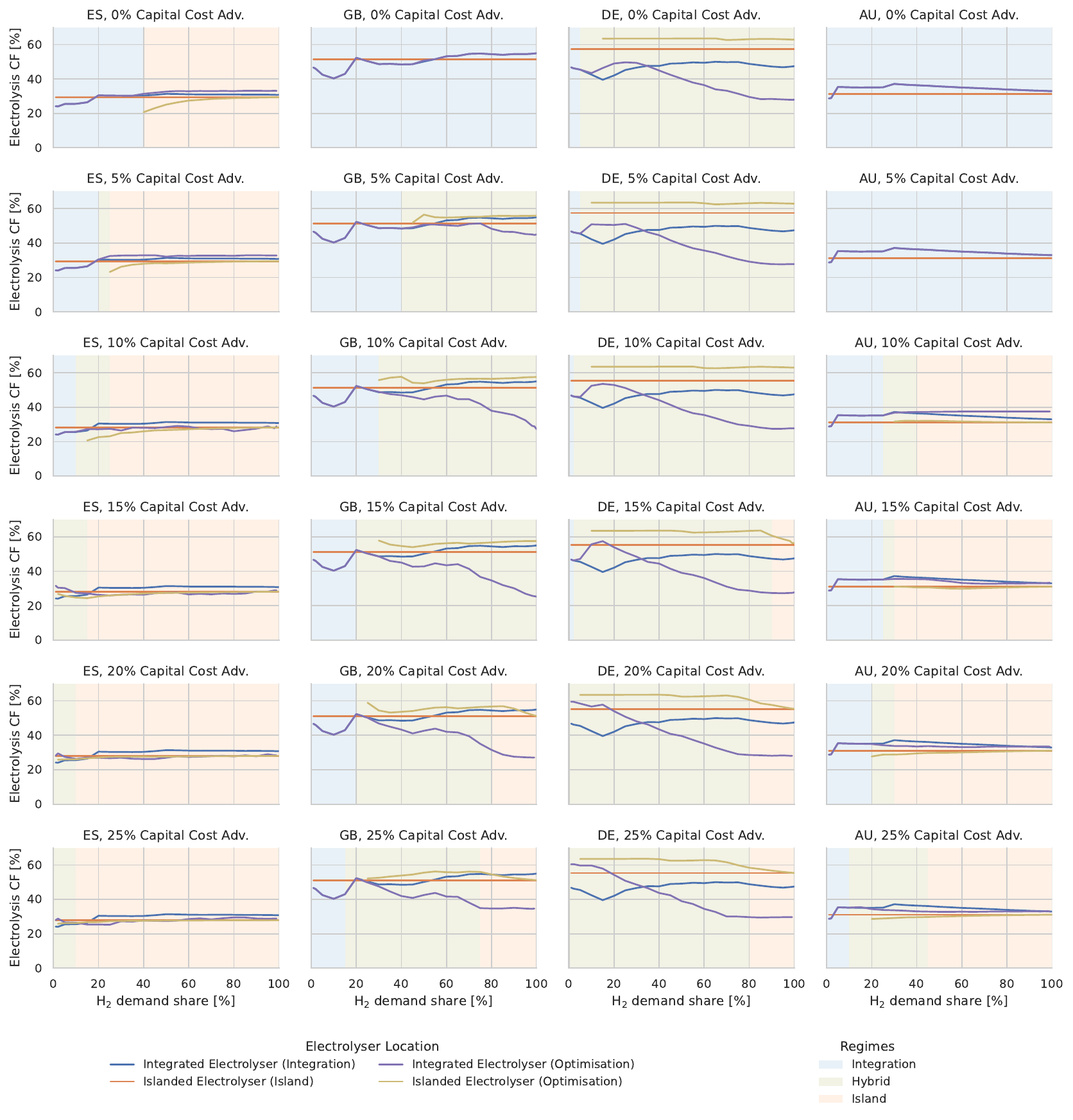}
  \caption{Electrolysis capacity factor for integrated and islanded electrolysers across countries and levels of capital cost reductions}
  \label{fig:cfElectrolysis-Grid}
\end{figure}

\begin{figure}[!ht]
  \centering
  \includegraphics[width=1.0\textwidth]{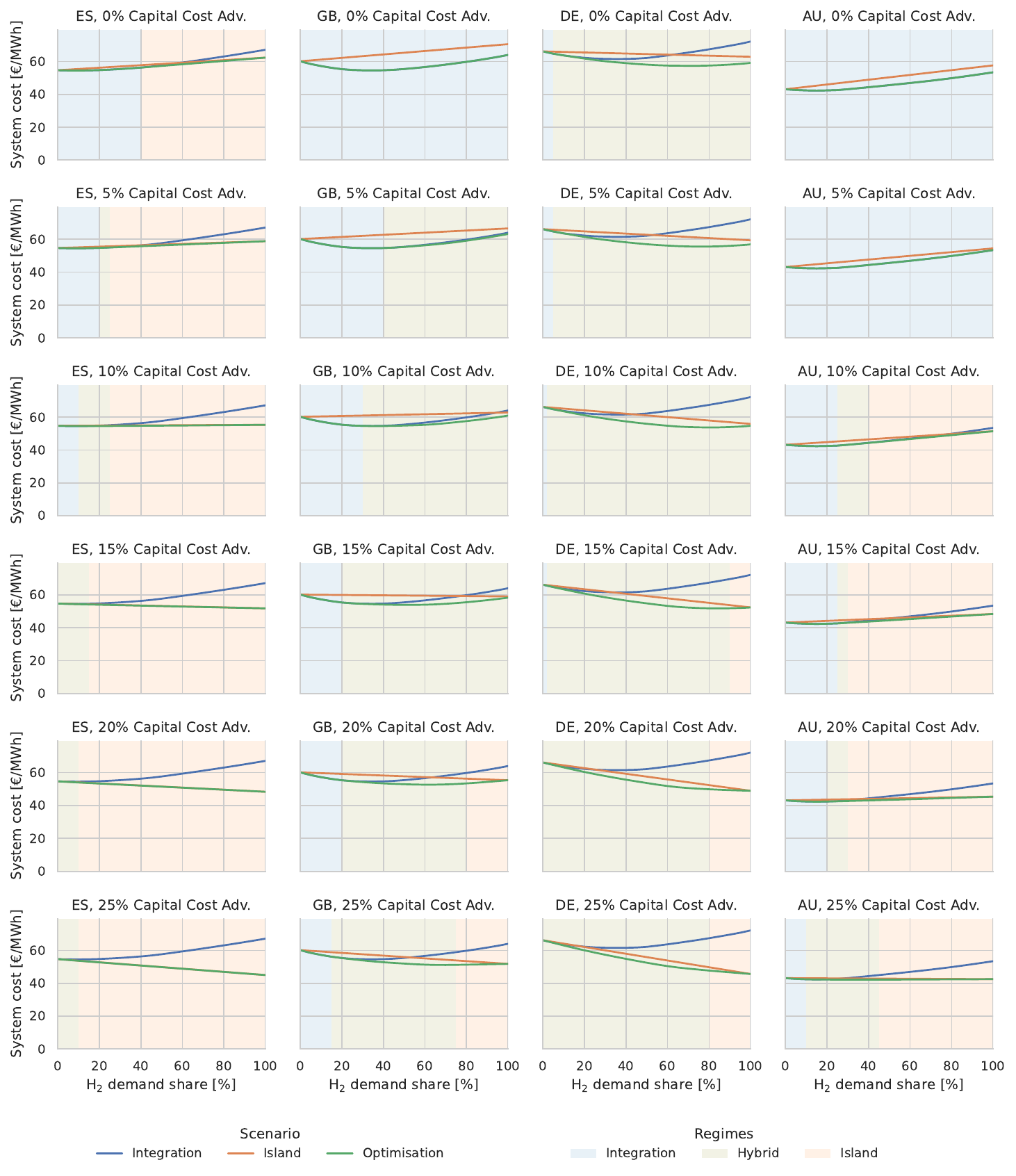}
  \caption{Per unit system costs across countries and levels of capital cost reductions}
  \label{fig:systemCosts-Grid}
\end{figure}

\end{document}